\documentclass{aa}

\usepackage{hyperref}
\usepackage{pdflscape}
\usepackage{xcolor}
\usepackage{lscape}

\usepackage{graphicx}
\usepackage[normalem]{ulem}
\newcommand{\kms}{km\,${\rm s}^{-1}$}
\newcommand{\Msun}{M$_{\odot}$}
\newcommand{\halpha}{H$\alpha$}
\newcommand{\Starname}{{HR\,6819}}
\newcommand{\vsini}{{$\varv\,\sin i$\,}}
\newcommand{\vmac}{$\varv_\mathrm{macro}$\,}
\newcommand{\vmic}{$\varv_\mathrm{micro}$\,}
\newcommand{\cgs}{[cgs]}
\usepackage{txfonts}
\begin{document}

   \title{Is HR 6819 a triple system containing a black hole?}
    \subtitle{An alternative explanation}

   \author{J. Bodensteiner, T. Shenar, L. Mahy, M. Fabry, P. Marchant, M. Abdul-Masih, G. Banyard, D. M. Bowman, \\ K. Dsilva, A. J. Frost, C. Hawcroft, M. Reggiani, H. Sana
   }

   \institute{Institute of Astronomy, KU Leuven, Celestijnenlaan 200D, B-3001 Leuven, Belgium \\
              \email{julia.bodensteiner@kuleuven.be}
    }

   \date{Received Month xx, xxxx; accepted Month xx, xxxx}

% \abstract{}{}{}{}{}
% 5 {} token are mandatory

  \abstract
  % context heading (optional)
   {
    \Starname{} was recently proposed to be a triple system consisting of an inner B-type giant + black hole (BH) binary with an orbital period of 40\,d and an outer Be tertiary. This interpretation is mainly based on two inferences: that the emission attributed to the outer Be star is stationary, and that the inner star, which is used as mass calibrator for the BH, is a B-type giant.
   }
  % aims heading (mandatory)
   {
   We re-investigate the properties of \Starname{} to search for a possibly simpler alternative explanation for \Starname{}, which does not invoke the presence of a triple system with a BH in the inner binary.
   }
  % methods heading (mandatory)
   {
  Based on an orbital analysis, the disentangling of the spectra of the two visible components, and the atmosphere analysis of the disentangled spectra, we investigate the configuration of the system and the nature of its components.
   }
  % results heading (mandatory)
   {
    Disentangling implies that the Be component is not a static tertiary, but rather a component of the binary in the 40-d orbit. The inferred radial velocity amplitudes of $K_1 = 60.4 \pm 1.0 \,$\kms{} for the B-type primary and $K_2 = 4.0 \pm 0.8\,$\kms{} for the Be-type secondary imply an extreme mass ratio of $M_2/M_1 = 15\pm3$. We find that the B-type primary, which we estimate to contribute about 45\% of the optical flux, has an effective temperature of $T_\mathrm{eff}=16\pm1$\,kK and a surface gravity of $\log g=2.8\pm0.2\,$\,\cgs, while the Be secondary, which contributes about 55\% to the optical flux, has $T_\mathrm{eff}=20\pm2$\,kK and $\log g=4.0\pm0.3$\,\cgs. We infer spectroscopic masses of $0.4^{+0.3}_{-0.1}$\,\Msun{} and $6^{+5}_{-3}$\,\Msun{} for the primary and secondary, which agree well with the dynamical masses for an inclination of $i = 32^\circ$. This indicates that the primary might be a stripped star rather than a B-type giant. Evolutionary modelling suggests that a possible progenitor system would be a tight ($P_{\rm i} \approx 2\,$d) B+B binary system that experienced conservative mass transfer. While the observed nitrogen enrichment of the primary conforms with the predictions of the evolutionary models, we find no indications for the predicted He enrichment. 
   }
  {
   We suggest that \Starname{} is a binary system consisting of a stripped B-type primary and a rapidly-rotating Be star that formed from a previous mass-transfer event. In the framework of this interpretation, \Starname{} does not contain a BH. Interferometry can distinguish between these two scenarios by providing an independent measurement of the separation between the visible components.
   %An explicit test to distinguish between the two scenarios and to reveal the configuration of the system is an independent measurement of the separation between the visible components, which can be provided by interferometry.
  } 

   % \titlerunning{xxx}
   \authorrunning{Bodensteiner et al.}

   \maketitle
%
%-------------------------------------------------------------------

%
\section{Introduction}\label{sec:intro}
Massive stars (stars with initial masses $\gtrsim8$\,\Msun{}) have a major impact on their immediate surroundings and the evolution of their entire host galaxies through their energetic radiation, their strong stellar winds, and especially their spectacular explosions \citep[e.g.,][]{Jamet2004, Bromm2009, Aoki2014}. Yet, a detailed understanding of their formation and complex evolution is still lacking \citep[see e.g.,][]{Langer2012}. 

One major uncertainty in their evolution is multiplicity. It has been shown observationally that the majority of massive stars are born in binary or higher-order multiple systems \citep{Sana2012, Kobulnicky2014, Dunstall2015, Moe2017}. A large fraction of these systems interact during their lives which drastically alters their evolutionary path and its outcome \citep{Paczynski1967, Podsiadlowski1992, Vanbeveren1994, deMink2013}.

Given the high rate of detected gravitational wave events that originate from the merger of two stellar-mass black holes \citep[BHs;][]{Abbott2019}, it is crucial to understand the origin of such stellar-mass BHs. Alongside dynamical \citep{DiCarlo2019, Fragione2019} as well as primordial \citep{Nishikawa2019} formation pathways, one of the proposed channels for double-BH formation is close binary evolution \citep[e.g.,][]{Belczynski2002, Mandel2016,  Marchant2016, Abdul-Masih2019}. Focusing on the binary channel, recent theoretical studies predict that a significant fraction of O- and early-B type stars are in binary systems with an unseen BH companion \citep{Langer2019, Shao2019, Yi2019}.

While a handful of these can be detected through X-ray emission in so-called high-mass X-ray binaries \citep[HMXBs, see e.g.,][]{Liu2006}, a majority of the systems could be X-ray quiet and are thus difficult to detect by observations. This is reflected by the lack of reported OB+BH binaries. Until recently, only one such X-ray quiet system, namely MWC 656, a Be+BH system \citep{Casares2014}, was known. Recently, two more similar systems were proposed to contain a stellar-mass BH: \object{LS V +22 25} (LB-1 hereafter) and \Starname{}.

\object{LB-1} was proposed by \cite{Liu2019} to contain a $\approx70$\,\Msun{} BH in a 79-d orbit with a B-type star companion. This interpretation relied on the adopted mass of the B-type primary and on the apparent low-amplitude motion of the \halpha{} line, which was interpreted to follow the orbit of the unseen secondary. 
Both these aspects were, however, challenged by subsequent studies. \cite{ElBadry2020} and \cite{AbdulMasih2020} showed that the anti-phase signature of \halpha{} cannot be readily interpreted as Doppler motion without considering the impact of potential H$\alpha$ absorption of the B-type component.
\cite{Simon-Diaz2020} and \cite{Irrgang2020} performed spectroscopic studies of the primary and found evidence for processed CNO material in its atmosphere, leading \citet{Irrgang2020} to suggest that it is a low-mass stripped helium star. 
Subsequently, \cite{Shenar2020} utilised spectral disentangling to show that the secondary in \object{LB-1} is not a BH, but  a classical Be star -- a rapidly rotating B-type star with a decretion disk \citep[see e.g.,][]{Rivinius2013}.
% Because of the strong broadening of its spectral lines, the companion was not readily seen in the individual observations, causing previous studies to assume that LB-1 hosts a compact object. 
% \citet{Shenar2020} demonstrated that the narrow-lined primary in \object{LB-1} indeed shows signatures of a stripped star, as proposed by \citet{Irrgang2020}, and that the spectrum of the secondary resembles the spectrum of a classical Be star -- a rapidly rotating B-type star with Balmer line emission forming in a decretion disk around the star \citep[see e.g.,][]{Rivinius2013}. 
% Calibrating the mass of the secondary to $\mathrm{M}_2 = 7 \pm 2$\,\Msun{}, \citet{Shenar2020} found a primary mass of $\mathrm{M}_1 = 1.5 \pm 0.4$\,\Msun{}. 
\citet{Shenar2020} interpreted LB-1 as a post-mass-transfer system \citep{Pols1991}, where the primary was stripped of most of its envelope and the secondary accreted mass and angular momentum to become a Be star. Based on an independent near-infrared study, \cite{Liu2020b} recently derived a mass ratio comparable to that reported by \citet{Shenar2020}, but remained inconclusive regarding the nature of the secondary. Hence, evidence for the presence of a BH in LB-1 is currently lacking.

The second system that was recently proposed to contain a stellar-mass BH is \Starname{} \citep{Rivinius2020}. It was reported to be variable already by \cite{Buscombe1960}. The composite nature of the spectrum consisting of a narrow-lined component and a broad-line component was reported by \cite{Hiltner1969} and \cite{Dachs1981}, who classified the system as B3~IIIep. A similar spectral classification, B3~III(e), was found by \cite{Slettebak1982},  who interpreted the star as a "normal sharp-lined B3 giant". \cite{Dachs1986} came to a similar conclusion and retracted their earlier interpretation of a composite spectrum. Based on FEROS data, \citet{Maintz2003} reported an emission profile in H$\beta$ moving in anti-phase with the narrow absorption, proposing that the system might contain a B+Be binary with a 30-60\,d period. \citet{Wang2018} studied the system in the UV and did not find any indications for a hot subdwarf companion.

Recently, \cite{Rivinius2020} suggested that \Starname{} hosts the closest stellar-mass BH to the Sun yet reported. %Based on an unpublished study (Hadrava et al.\ in prep.),
They argued that the emission profiles attributed to the rapidly-rotating Be star are stationary on the time scale of years, and that the Be star seen in the spectrum must therefore be a distant tertiary in an outer orbit. Based on the binary mass function ($f_{\rm M} = 0.96$\,\Msun{}) derived from the SB1 RV curve and the mass estimate of the narrow-lined star ($M \ge 5\,$\Msun{}), which the authors classified as B3~III, \cite{Rivinius2020} concluded that the unseen secondary star in the 40-d orbit must be more massive than $4\,$\Msun{}. Given this mass estimate, they concluded that the unseen companion must be a BH. The system was thus interpreted by \citet{Rivinius2020} as a hierarchical triple containing a B + BH inner binary on a 40-d orbit and a Be star on an unconstrained wider orbit. They suggested a similar configuration for LB-1 which is, however, contradicted by \cite{Shenar2020} and \cite{Liu2020b}.
% PhD thesis of Maintz: http://archiv.ub.uni-heidelberg.de/volltextserver/4343/1/Monika_Maintz_Dissertation.pdf

% Rotational velocities low: 50+/-5 Slettebak 1982
% 47+/-6 https://ui.adsabs.harvard.edu/abs/1997A%26A...319..811B/abstract
% 48+/-7 https://ui.adsabs.harvard.edu/abs/2001A%26A...378..861C/abstract
% 66+/-8 Yudin+2001
% 48 Fremat+ 2005

% https://ui.adsabs.harvard.edu/abs/1989MNRAS.241..721P/abstract:  T$_\mathrm{eff}$ = 150000K, log L/Lsun = 3.62, r/rsun = 5.1, M/Msun = 9, vsini = 50, vesc=840

% no detection in the rosat all-sky catalogue https://ui.adsabs.harvard.edu/abs/1996A%26AS..118..481B/abstract

% listed as probably single in Eggleton+ 2008

% not listed as runaway in Tetzlaff+2011

% Chini+ 2012: SB2

% Zorec+ 2016: T$_\mathrm{eff}$ ~ 20000 +/- 600, $\log g$ ~ 3.5+/- 0.3, logL/Lsun ~ 3.4, E(B-V) ~ 0.135, sini 48+/-4

% BeSOS survey 2018: T$_\mathrm{eff}$ ~ 20000+/- 200, $\log g$~3.8+/-0.04, R/Rsun ~3.9+/-0.08, E(B-V)~0.14, vsini = 50 +/- 1 km/s ( https://ui.adsabs.harvard.edu/abs/2018MNRAS.474.5287A/abstract) 

In this paper we present an alternative explanation to \Starname{} that does not invoke an inferred stellar-mass BH or a tertiary component on a wider orbit. Our study is based on the same observational data as used by \citet{Rivinius2020}, which is described in Sect.\,\ref{sec:obs}. In Sect.\,\ref{sec:variability} we derive the orbit of the narrow-lined primary and show additional sources of variability in the composite spectra. In Sect.\,\ref{sec:disen} we describe the disentangling of the spectra while we derive stellar parameters in Sect.\,\ref{sec:specclass}. To investigate the origin and distribution of emission in the system we apply the Doppler tomography technique in Sect.\,\ref{sec:tomography}. A possible evolutionary history of the system is discussed in Sect.\,\ref{sec:discussion}, while the conclusions are given in Sect.\,\ref{sec:concl}.

\section{Observations and data reduction}\label{sec:obs}
Two archival data sets of \Starname{} are available, both obtained with the Fibre-fed Extended Range Optical Spectrograph \citep[FEROS,][]{Kaufer1999}. The first data set, acquired in 1999, contains 12 epochs\footnote{program ID 63.H-0080, available under https://www.lsw.uni-heidelberg.de/projects/ instrumentation/Feros/ferosDB/search.html} while the second data set from 2004 contains 51 epochs\footnote{program ID 073.D-0274, available on the ESO archive under http://www.eso.org/sci/ facilities/lasilla/instruments/feros.html}.

Between the two independent data sets, FEROS was moved from the ESO 1.52-m telescope to the 2.2-m MPG/ESO telescope both situated in the La Silla Observatory, Chile. FEROS covers a wavelength range between 3500 and 9200\,$\AA$ with a spectral resolving power of 48\,000. The journal of observations, including typical signal-to-noise ratios (S/N) and radial velocities (RVs) measured as described in Sect.\,\ref{subsec:orb_analysis} is given in Table\,\ref{tab:obs_log}.

While observations of the first data set were downloaded fully-reduced, the second data set was reduced using the standard FEROS pipeline (version 1.60). Both include all standard calibrations such as bias and flat-field corrections and wavelength calibrations. A barycentric correction was applied to all spectra. The homogeneity of the wavelength calibration between the two data sets was reviewed using the two interstellar Na\,\textsc{i} lines at 5890 and 5896 $\AA$. All 63 fully-reduced spectra were individually normalized by fitting a spline to selected anchor points in the continuum of each spectrum.

In addition to the aforementioned FEROS observations, we also consider spectra available in the BeSS\footnote{\textsc{http://basebe.obspm.fr/basebe/}} database \citep{Neiner2011}. These spectra cover a longer time span (i.e. approximately one observation per year since 2011, see Table\,\ref{tab:obs_log_bess}), %Given that they are mostly taken by amateur astronomers,
however, their spectral coverage and quality varies. We thus only use them in a qualitative sense in order to investigate variability in the \halpha{} line (see Sect.\,\ref{subsec:halpha_variability}).

\begin{table}
\centering
    \caption{ \textit{Upper}: Derived orbital parameters based on our orbital analysis and disentangling, along with their 1$\sigma$ errors. \textit{Lower}: Estimated physical parameters of the components of \Starname{} based on a spectral analysis with PoWR and TLUSTY. The luminosity and radii are obtained for an assumed distance of $340\,$pc \citep{Bailer-Jones2018}.}
    \begin{tabular}{llcc} \hline \hline
        Parameter & primary & secondary \\ 
        \hline
        Spectral type    & stripped star & B2-3~Ve  \\     
         $P_\mathrm{orb}$\,[d] & \multicolumn{2}{c}{$40.335 \pm 0.007$} \\
         $T_0$\,[MJD] ($\Phi=0$) & \multicolumn{2}{c}{$53116.9 \pm 1.1$} \\
         $e$  & \multicolumn{2}{c}{0 (fixed)} \\
         $\gamma$\,[\kms{}] &  \multicolumn{2}{c}{9.13 $\pm$ 0.78} \\
         $K$\,[\kms{}] &  60.4 $\pm$ 1.0 & $4.0 \pm 0.8$ \\
         $M\,\sin^3 i$\,[$M_\odot$] &  $0.07 \pm 0.03$ & $1.05 \pm 0.02$ \\
         $a\,\sin i$\,[$R_\odot$] &  $48.2 \pm 0.8$ & $3.2\ \pm 0.6$ \\      
         $q(\frac{M_2}{M_1})$  &   \multicolumn{2}{c}{$15 \pm 3$} \\
         $M_{\rm dyn}$\,[$M_\odot$] &  $0.46 \pm 0.26$ & $7\pm2$ (fixed) \\   
         $i$\,[deg] &  \multicolumn{2}{c}{32$ \pm $4}\\
         \hline
        $T_{\rm eff}$\,[kK] & $16\pm1$ & $20\pm2$   \\
        $\log g$\,\cgs & $2.8\pm0.2$ & $4.0\pm0.3$   \\
        flux $f/f_{\rm tot}(V)$ & 0.45 (fixed)  & 0.55 (fixed)  \\
        $\log\,L\,[L_\odot]$  & $3.05\pm0.10$  & $3.35\pm0.10$  \\  % 3.2
        $R\,[R_\odot]$ & $4.4 \pm 0.4$ & $3.9\pm0.7$  \\ % 5.9 
        $M_{\rm spec}$\,[$M_\odot$] & $0.4_{-0.1}^{+0.3}$  & $6^{+5}_{-3}$  \\  % 0.5
        \vsini{}[\kms{}] & $\lesssim 25$ & $180\pm10$  \\
        $\varv_{\rm eq}$\,[\kms{}] & $\lesssim 40$ & $340\pm40$ \\        
        \vmac{}[\kms{}] & $35\pm5$ & $70\pm25$  \\           
        \vmic{}[\kms{}] & 10 (fixed) & 2 (fixed) \\
        % $Z/Z_\odot$ &  1  & 1 \\ 
        % n(He)/n(H)  & 0.21 & solar (0.08) \\
    \hline
    \end{tabular}
    \label{tab:Bstar_properties}
\end{table}

\section{Spectral variability}\label{sec:variability}

\Starname{} exhibits spectral variability on the order of days, weeks, and years. The most apparent form of variability is seen through the RV shifts of the narrow-lined primary, which follows a Keplerian orbit of about 40\,d. However, other forms of variability, manifested through changes in the line profiles, widths, and depths, are also seen. 

\subsection{Orbital analysis} \label{subsec:orb_analysis}
We measure RVs of the narrow-lined primary star in the combined data set by fitting Gaussian profiles to several spectral lines simultaneously at all observing epochs, based on the method described in \cite{Sana2013}. Here, we use several He\,\textsc{i} and metal lines (i.e., N\,\textsc{ii}, Si\,\textsc{ii}, Si\,\textsc{iii}). We fit an orbit to the RV data using the python package \textsc{spinOS} (Fabry et al. in prep.)\footnote{\textsc{https://github.com/roeiboot4/spinOS}}.

We find that the root-mean-square (rms) of the orbit is 4.5\,\kms{}, which is larger than the typical formal RV errors. Since the formal RV errors are underestimated given the variability in the spectral lines (see Sect.\,\ref{subsec:HeI} and \ref{subsec:Si}), we scale the RV errors by the ratio between the rms and the mean error. The derived eccentricity is consistent with zero within the errors and we thus conclude the orbit to be circular, with $\omega$ fixed to $90^{\circ}$. Accordingly, T$_0$ corresponds to the primary conjunction. Our orbital parameters (see Table \,\ref{tab:Bstar_properties}) agree well with the parameters reported by \cite{Rivinius2020}.

\subsection{Apparent anti-phase motion of the \halpha{} line} \label{subsec:Michael_effect}

% As discussed by \citet{Rivinius2020}, the composite spectra of \Starname{} bear a strong resemblance to those of \object{LB-1}. In both systems, an RV-variable narrow-lined component can be readily seen, along with double-peaked Balmer emission profiles characteristic of Be stars. However, unlike \object{LB-1}, the presence of broad absorption features primarily belonging to  He\,{\sc i} lines is also readily apparent in \Starname{}, owing to the high S/N of the data. For \object{LB-1}, \citet{Shenar2020} concluded that the spectral lines of the Be companion move in anti-phase with the stripped primary star. The question therefore arises whether this is the case for \Starname{} as well, implying that the Be component in \Starname{} is the companion of the narrow-lined star, and thus questioning the presence of a BH in the \Starname{} system.

\begin{figure}
\centering
\includegraphics[width=0.5\textwidth]{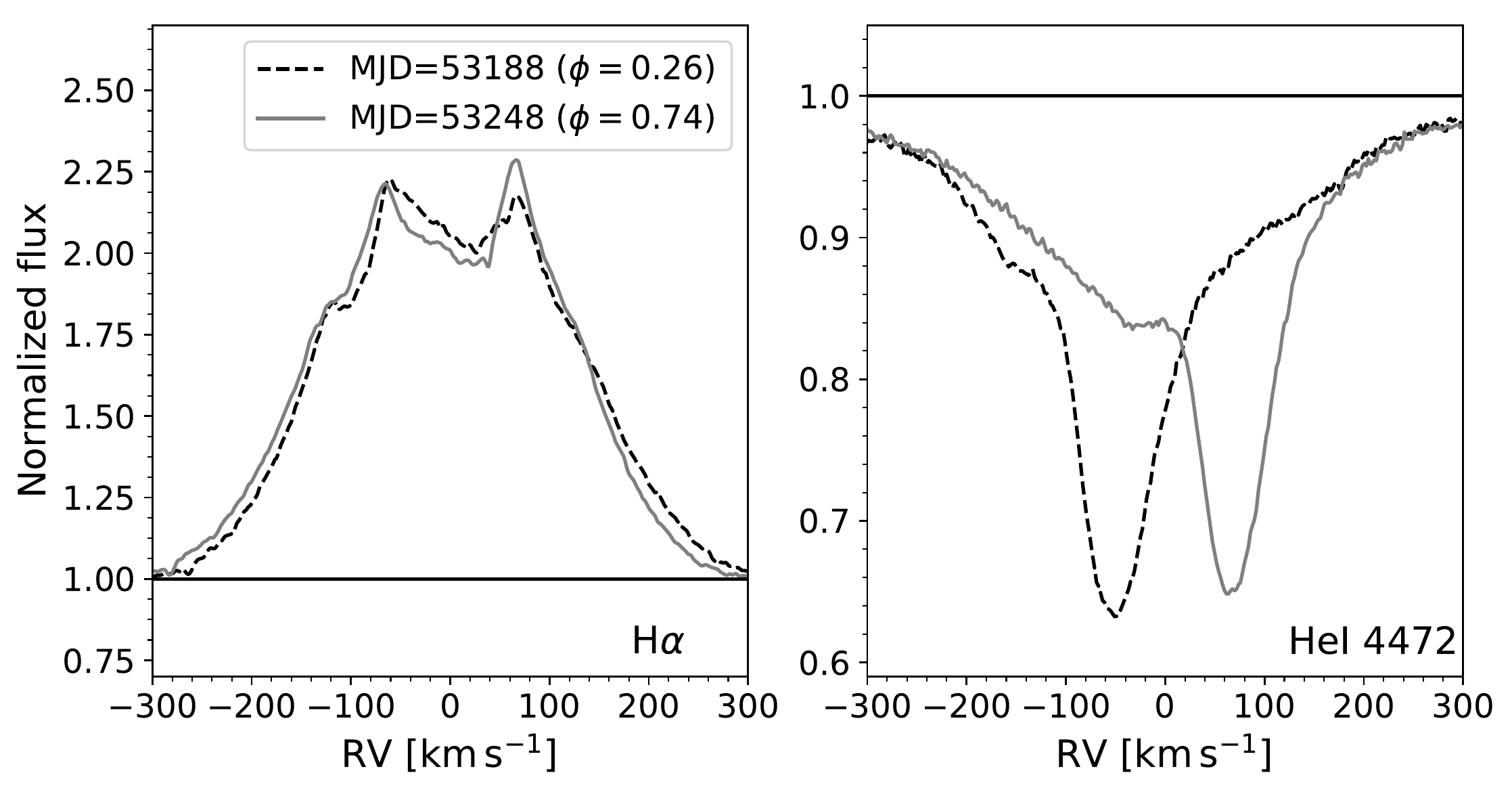}
\caption{Two FEROS spectra taken close to quadrature, focusing on the \halpha{} (left) and He\,{\sc i}\,$\lambda 4472$ (right) lines. We note the apparent anti-phase motion observed in the \halpha{} line.}
\label{fig:HalphaMoving}
\end{figure}

% Just as for \object{LB-1},
The \halpha{} line in \Starname{} appears to exhibit an anti-phase motion with respect to the primary. This is illustrated in Fig.\,\ref{fig:HalphaMoving}, where two FEROS spectra taken close to quadrature (maximum RV shift) are shown. However, it is not clear without further analysis whether this behaviour is the result of the orbital motion of the Be component, or the superposition of an absorption component in \halpha{} stemming from the narrow-lined primary \citep{AbdulMasih2020, ElBadry2020}. % It is also interesting to note that

The core of the \halpha{} line remains relatively unchanged. This implies that the narrow-lined primary does not portray significant absorption in \halpha{}.
Instead, \Starname{} exhibits a behaviour similar to other Be binaries such as \object{$\pi$\,Aqr} \citep{Bjorkman2002} and \object{o\,Pup} \citep{Koubsky2012}, % and \object{LB-1} \citep{Shenar2020},
where a trailing emission in \halpha{} is reported to follow the orbit of the low-mass object in the system. The lack of significant absorption in the narrow-lined component may suggest that the apparent anti-phase RV shift observed in \halpha{} is the result of an actual orbital motion of the Be component.

\subsection{Short-term variability of He\,{\sc i} lines}\label{subsec:HeI}

\begin{figure}
\centering
\includegraphics[width=0.5\textwidth]{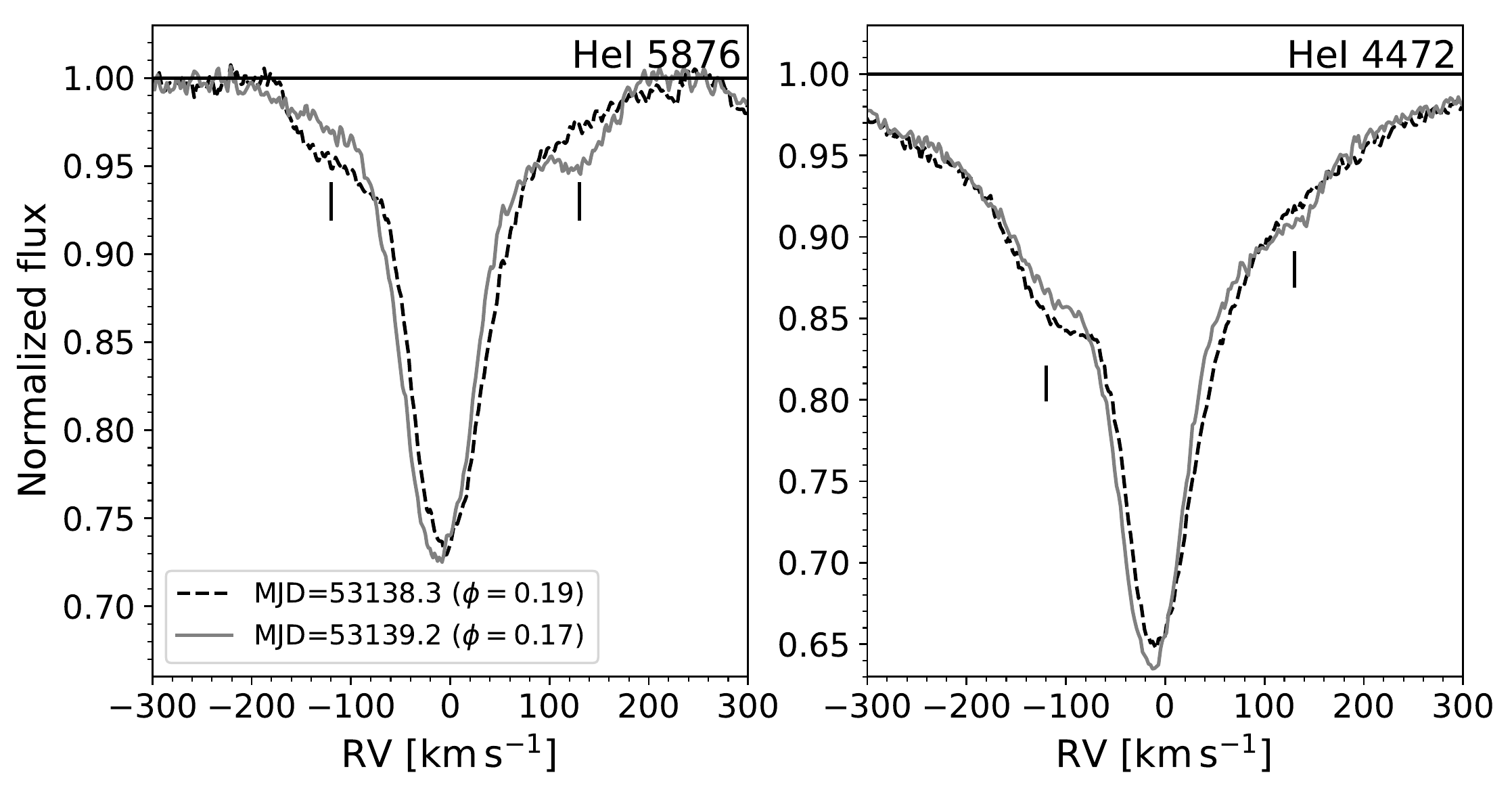}
\caption{Two FEROS observations taken on consecutive days, focusing on two He\,{\sc i} lines (see legend and labels). We mark the strong variability features seen in the wings of the He\,{\sc i} lines. }
\label{fig:HeI_Var}
\end{figure}

Significant variability is observed in strong He\,{\sc i} lines. Figure\,\ref{fig:HeI_Var} shows two FEROS observations of the He\,{\sc i}\, $\lambda\,4472$ and $\lambda\,5876$ lines taken approximately 1~d apart from each other. Very strong variability is apparent in the wings of the He\,{\sc i}\, $\lambda\,5876$ line and, to a lesser extent, the He\,{\sc i}\, $\lambda\,4472$ line. The data appear as if two additional stellar components move rapidly from one side to the other on a very short orbit (i.e., of the order of a day). Such variability can, however, also be explained by non-radial gravity-mode pulsations. This is observed for other Be stars \citep[e.g.,][]{Baade1984, Rivinius2003, Papics2017} and demonstrated by simulations of line profile variability caused by gravity-mode pulsations \citep{Aerts2009}. 

We test the hypothesis of two additional components on an even shorter orbit in the \Starname{} system through multiple-Gaussian fits and period searches. Based on the currently available data, which are typically taken with at least one night in between, we can neither derive a conclusive solution for such a configuration, nor disprove it. In the rest of our analysis, we therefore assume that the system consists of two visible components, and that the variability observed in the strong He\,{\sc i} lines is intrinsic to the Be star. This is supported by previous time series spectroscopic studies of Be stars that demonstrate that He\,{\sc i} spectral lines are particularly sensitive to intrinsic pulsational variability in moderate and fast-rotating early-B stars \citep{Balona1999, Rivinius2003, Aerts2010}. 
We cannot, however, fully rule out the presence of an inner short-period binary in this 40-d system. Testing this would require securing time-series spectroscopy over the course of one night to sample the short-period (< 1~d) variability.

\subsection{Medium-term variability of the narrow-lined primary}\label{subsec:Si}

\begin{figure}
\centering
\includegraphics[width=0.5\textwidth]{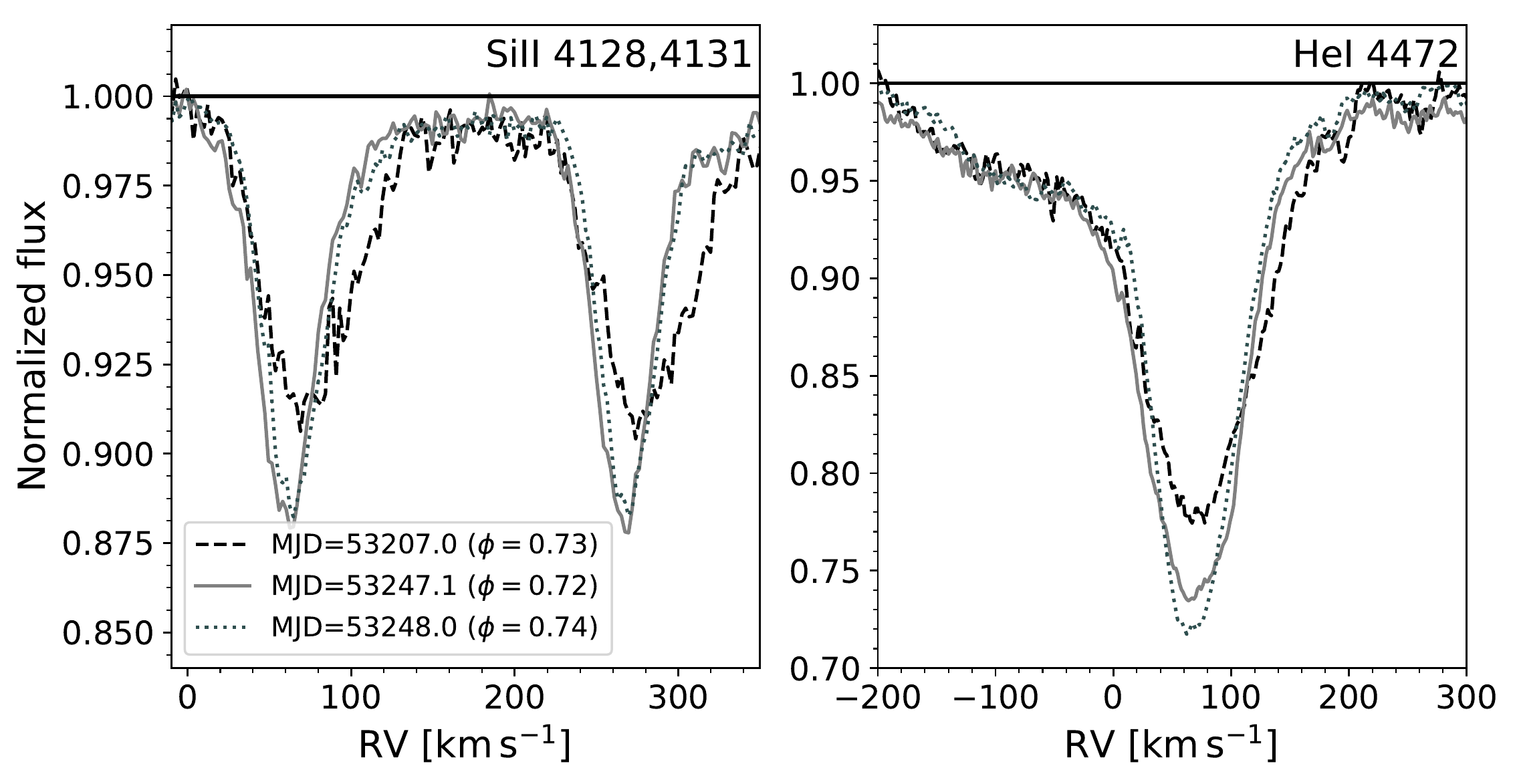}
\caption{Three FEROS observations taken at $\phi\approx 0.73$ but at three different days with a time separation of about 1\,d and 40\,d (see legend). The panels zoom in on two Si\,{\sc ii} lines, dominated by the narrow-lined primary, and on the He\,{\sc i} line, illustrating the apparent variability of the primary's line profiles on the scale of weeks.}
\label{fig:Si_Var}
\end{figure}

Significant variability is also observed in the spectrum of the narrow-lined primary. Figure\,\ref{fig:Si_Var} shows three FEROS observations focusing on the Si\,{\sc ii} $\lambda\lambda\,4128,\,4131$ doublet and the He\,{\sc i}\,$\lambda\,4472$ line, taken at approximately the same orbital phase, but with relative separations of about 1\,d and 40\,d. The Si\,{\sc ii} lines are, as all other metal lines, strongly dominated by the narrow-lined primary. As illustrated in Fig.\,\ref{fig:Si_Var}, the line width and depth change significantly over the course of a few weeks, but remain fairly similar over the course of days. 

The variability in the Si doublet can also be explained by non-radial gravity-mode pulsations \citep[see ][their figure 1]{Aerts2009}, with the silicon doublet (and triplet) being particularly sensitive to pulsations in slowly-rotating stars \citep{Aerts2009, Aerts2010}. Hence, we also interpret this as evidence for non-radial gravity-mode pulsations in the narrow-lined primary. The presence of pulsations is also related to the relatively large macroturbulence derived \citep[][see Sect.\,\ref{subsubsec:linebroad}]{Simon-Diaz2017}, since stars with larger pulsation amplitudes typically show larger macroturbulence \citep{Bowman2020}. We also note that \cite{Rivinius2020} demonstrate variability caused by non-radial gravity-mode pulsations using SMEI and TESS photometry, with such low-frequency photometric variability being common among massive stars \citep{Bowman2019b}.

\subsection{Long-term variability of Balmer lines}\label{subsec:halpha_variability}
Long-term variability in Balmer lines is a common feature of Be stars \citep{Lacy1977, Rivinius2013}. Different types of variability, probably caused by variations in the disk properties, are known to affect both the shape as well as the strength of the \halpha{} line. In double-peaked emission lines of some Be stars, so-called violet-to-red (V/R) cycles occur in which the strength of the two emission peaks varies cyclically with respect to each other \citep{Okazaki1991, Stefl2009}. On the other hand the overall strength of \halpha{} varies on the timescales of years to decades, where the emission can also fully disappear. This probably coincides with the formation and depletion of the circumstellar disk itself \citep{Doazan1986, Wisniewski2010, Draper2011}.

\begin{figure}
\centering
\includegraphics[width=0.5\textwidth]{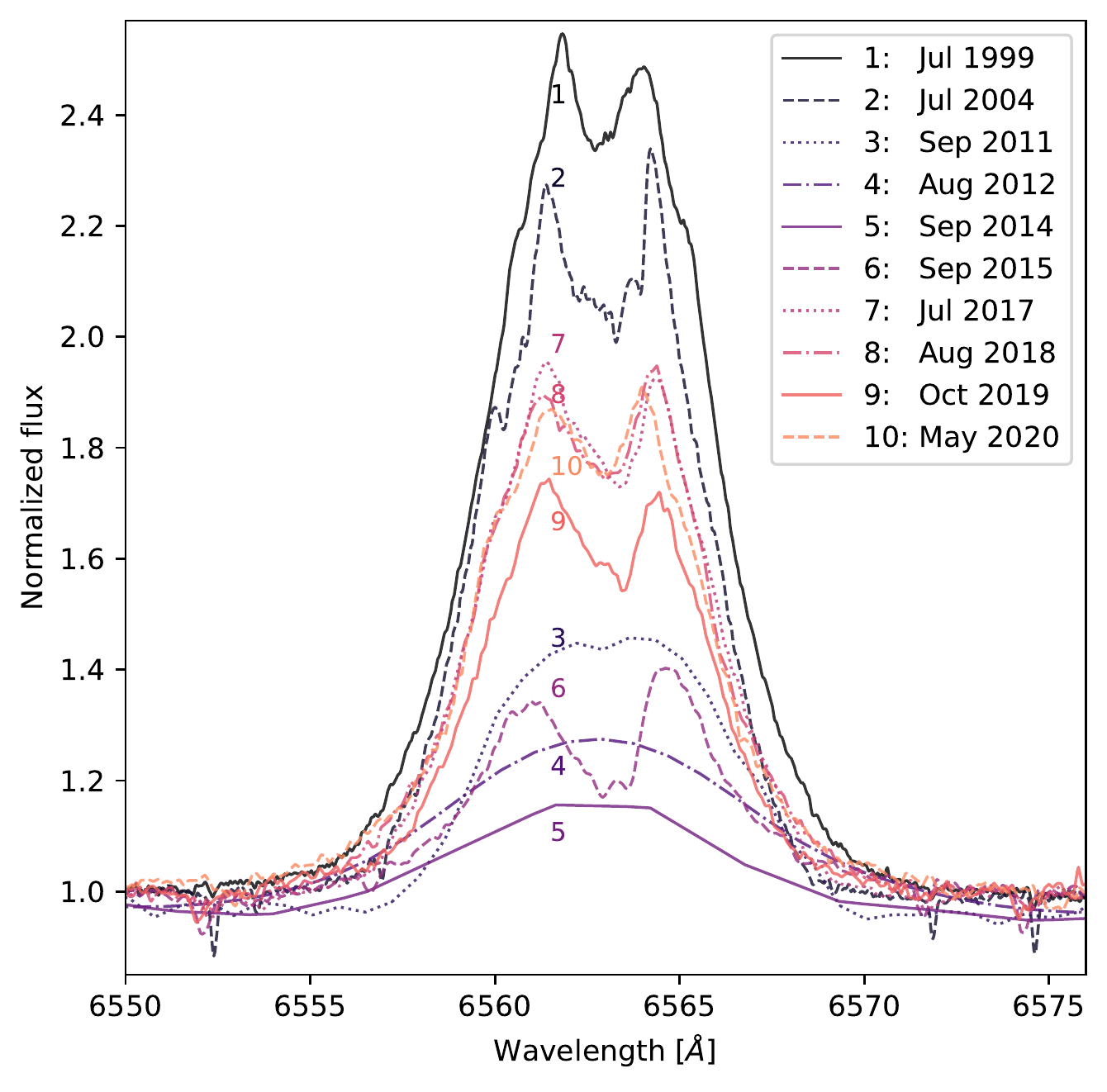}
\caption{Evolution of the \halpha{} line in the past 20 years. In addition to the BeSS spectra (more information is given in Table\,\ref{tab:obs_log_bess}) we plot one spectrum from each FEROS run (i.e., 1999 and 2004, see legend).}
\label{fig:BeSS}
\end{figure}

We investigate the FEROS and BeSS spectra of \Starname{} that cover different time scales. Figure\,\ref{fig:BeSS} shows that the strength of the \halpha{} line was highly variable over the past 20 years, while its shape remains rather similar (considering the different resolving powers).
% The variability of the \halpha{} line would have a strong impact on spectral disentangling (see Sect.\,\ref{sec:disen}), which is {\bf one of the reasons for not considering} the \halpha{} line in our final measurement of the semi-amplitude $K_2$.
The other Balmer lines portray a similar variability, albeit with a much weaker magnitude.
%Given that the line strength is stable on time scales of months (indicated by the 1999 and 2004 epochs separately), the impact of this variability on the disentangling presented in Sect.\,\ref{sec:disen} is expected to be limited.

\section{Spectral disentangling}\label{sec:disen}

We perform spectral disentangling to separate the observed composite spectra into the intrinsic spectra of the stellar components in \Starname{}, and to investigate whether they are bound in the 40-d orbit. We use two independent methods, the shift-and-add and the Fourier technique, described below.
% The disentangling procedure 
Both disentangling procedures depend on the orbital parameters of the system.
Having constrained all orbital parameters but $K_2$, we perform spectral disentangling along the $K_2$-axis. Specifically, we adopt the orbital parameters derived in Sect.\,\ref{subsec:orb_analysis} and perform multiple disentangling procedures varying $K_2$ from 0 to 20\,\kms{} in steps of 0.5\,\kms{}. In each step we compute the $\chi^2$ of the solution by comparing the disentangled spectra with the observed ones at all available phases. % We also test our results creating artificial data sets and applying both disentangling methods to them (see Appendix\,\ref{sec:simulations}).

When disentangling the spectra of \Starname{}, we thus implicitly assume that the Be component participates in the 40-d orbit. If true, our method should retrieve a non-vanishing value for the orbital semi-amplitude $K_2$. Otherwise, values consistent with zero should be obtained, implying that the Be component is a tertiary star.

Owing to the high S/N in the spectra of \Starname{}, a multitude of lines can be probed independently. We thus perform disentangling on a series of lines belonging to He\,{\sc i} (photospheric absorption), Fe\,{\sc ii} and O\,{\sc i} (disk emission), and Balmer lines (combination of photospheric absorption and disk emission). Given the large variability observed in the H$\alpha$ line (see Sect.\,\ref{subsec:halpha_variability}), we do not consider H$\alpha$ in our analysis. 

Because of the strong variability between the 1999 and 2004 epochs (see Sect.\,\ref{subsec:halpha_variability}), we furthermore refrain from combining both data sets. Instead, we focus on the 2004 epoch, for which 51 spectra are available that cover approximately four orbital cycles. A comparison with the 1999 epoch is given in Appendix\,\ref{sec:1999}.

The shift-and-add technique \citep[e.g.,][]{Marchenko1998, Gonzalez2006, Mahy2012, Shenar2018} is an iterative procedure, in which the disentangled spectra calculated in the $j^\mathrm{th}$ iteration, $A_j$ and $B_j$, are used to calculate the disentangled spectra for the $j+1^\mathrm{th}$ iteration.
%Typically, convergence is obtained after a few tens to hundreds of iterations, depending on the spectral morphology and orbital parameters.
To avoid spurious features that are typically observed in the wings of broad lines when using the shift-and-add method \citep[e.g.,][]{Marchenko1998}, we force the solution to be smaller than unity (within the S/N) in regions where it is not expected to exceed unity.

\begin{figure}
\centering
\includegraphics[width=0.5\textwidth]{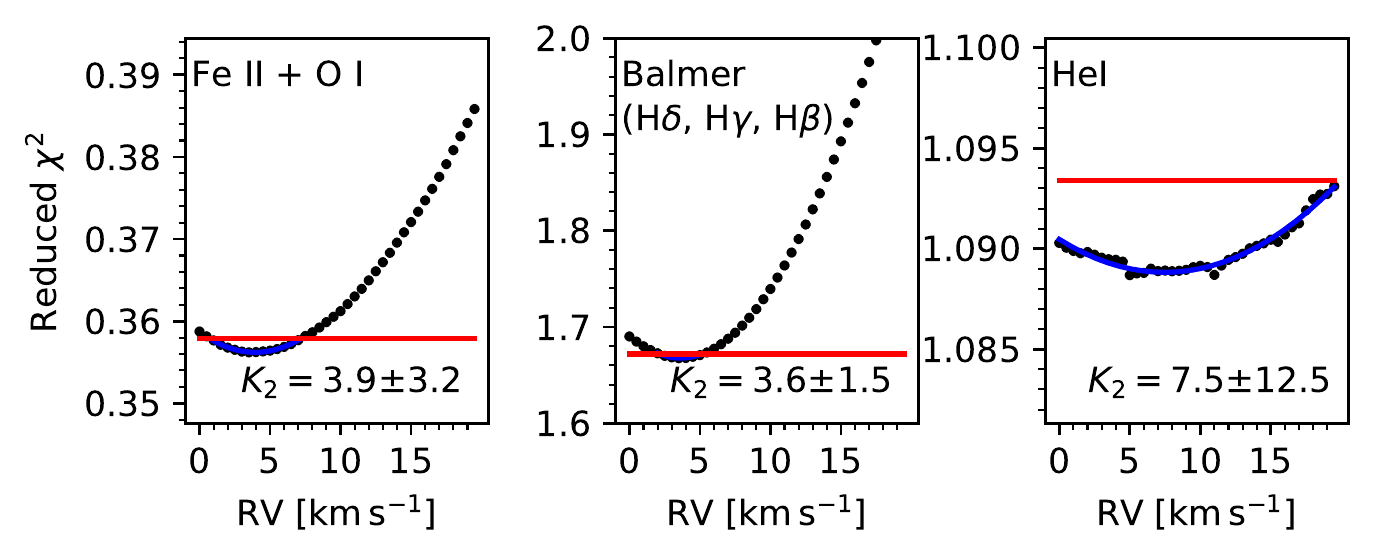}
\caption{Reduced $\chi^2(K_2)$ and corresponding $K_2$ measurements for individual line groups. The red lines depict the $1\sigma$ confidence interval. }
\label{fig:MasterGridDis_Main}
\end{figure}

\begin{figure}
\centering
\begin{tabular}{c}
     \includegraphics[width=0.5\textwidth]{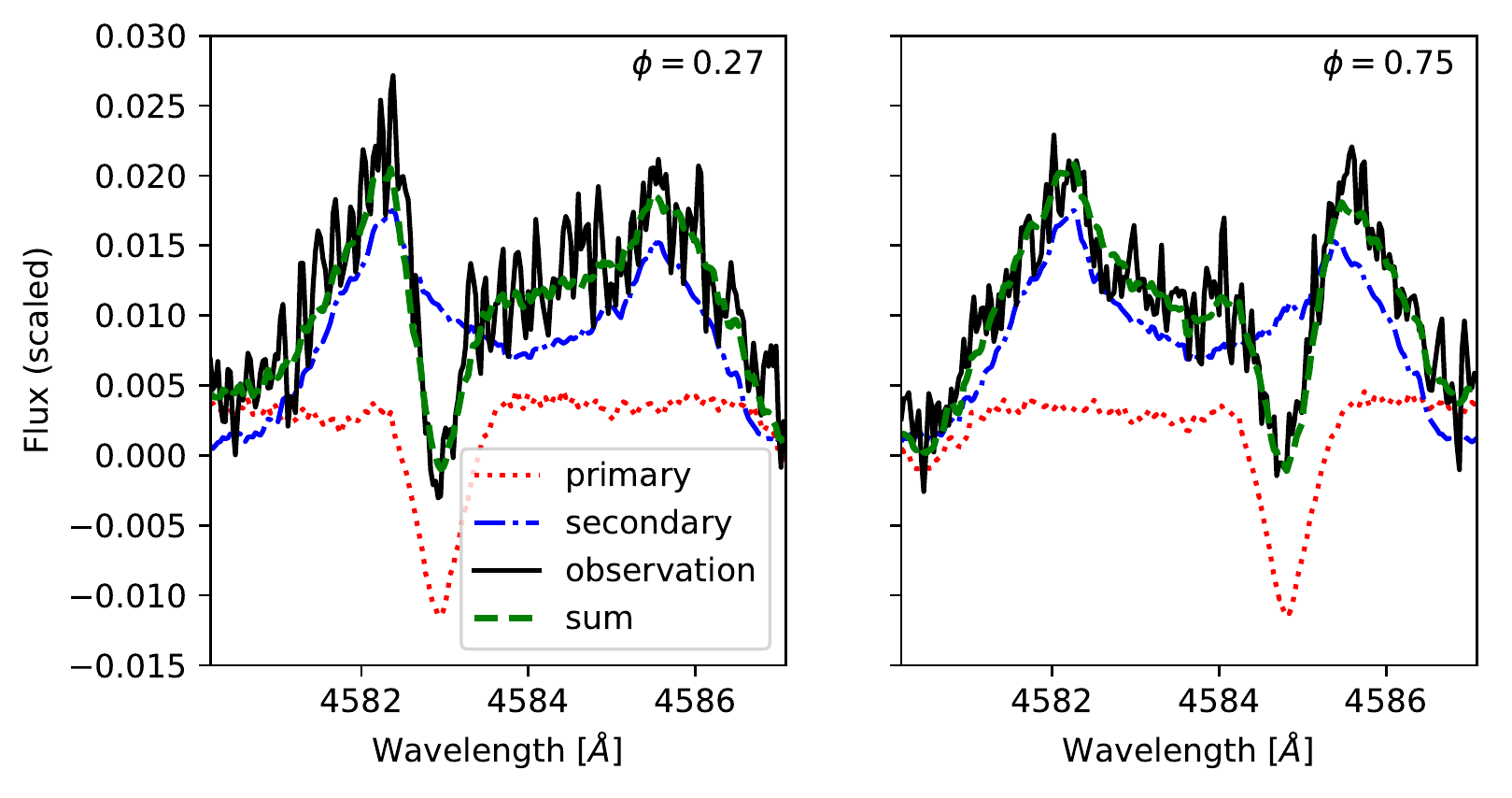} \\
     \includegraphics[width=0.5\textwidth]{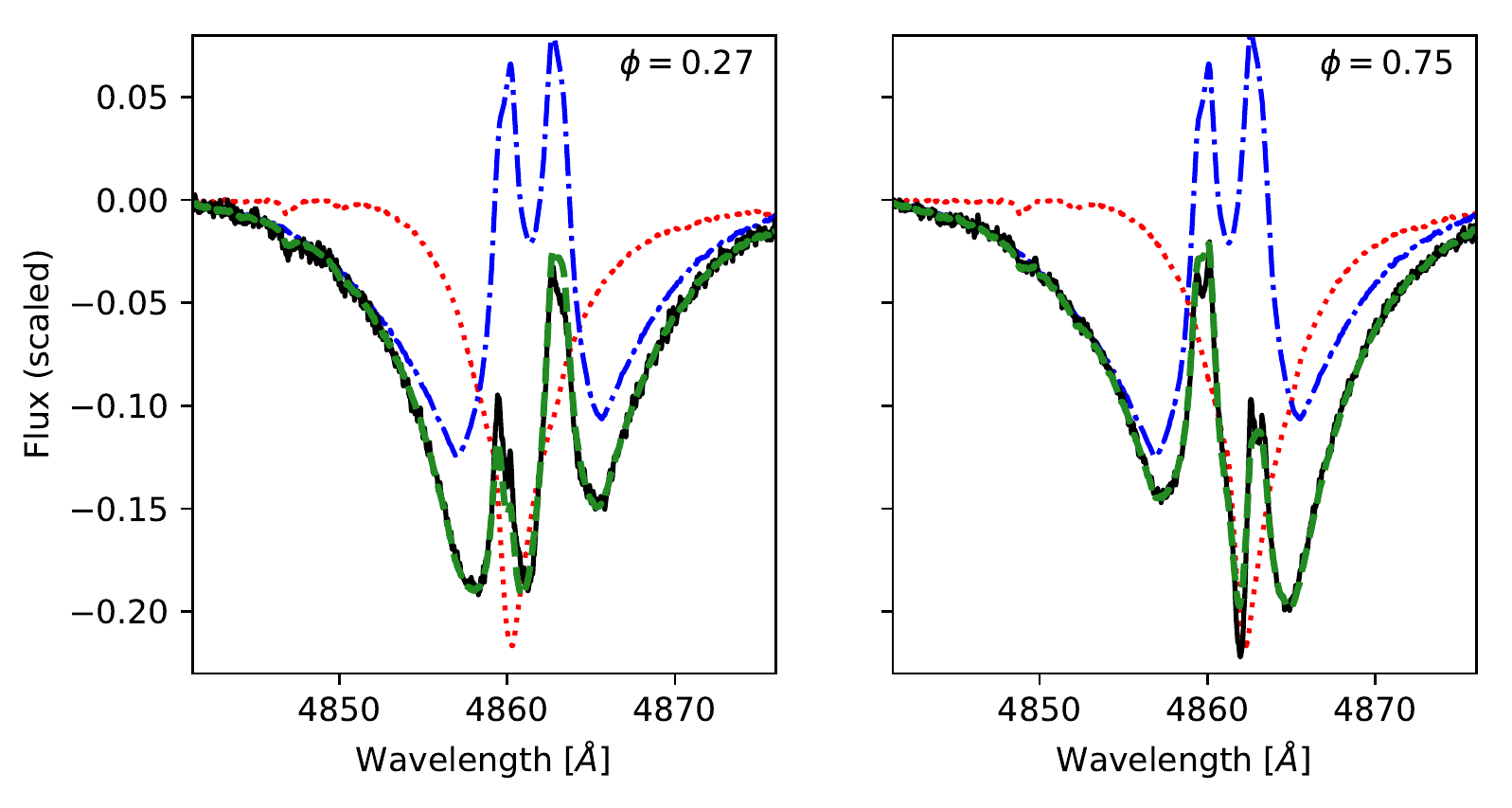} \\
     \includegraphics[width=0.5\textwidth]{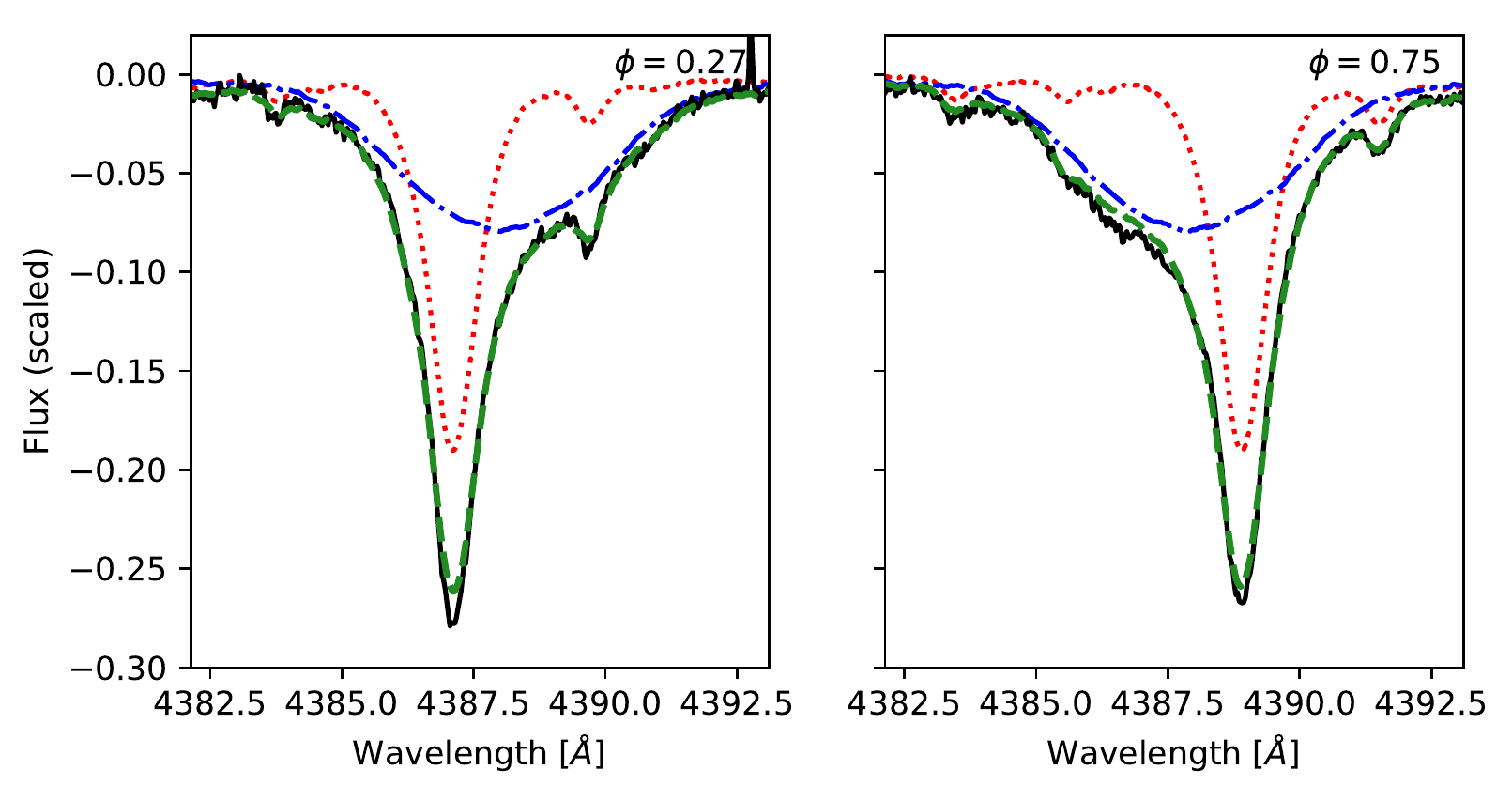}
\end{tabular}
\caption{Illustration of the fit quality between the individual observations at quadrature ($\phi = 0.25$ and 0.75 for the left and right columns, respectively) and the shift-and-added disentangled spectra. Fits are shown for the Fe\,{\sc ii}\,$\lambda 4584$ line (top), H$\beta$ line (middle), and He\,{\sc i}\,$\lambda 4388$ line (bottom). }
\label{fig:Quality}
\end{figure}

The results of the disentangling using shift-and-add are given in Fig.\,\ref{fig:MasterGridDis_Main}, where we show reduced $\chi^2(K_2)$ and corresponding $K_2$ measurements for the different line groups used as well as the best-fit. We note that the reduced $\chi^2$ obtained is of the order of unity for all lines considered (see Fig.\,\ref{fig:MasterGridDis_Main}). $K_2$ measurements of the individual lines are provided in Table\,\ref{tab:K2Meas}. Examples for the fits between individual observations and the disentangled spectra are shown in Fig.\,\ref{fig:Quality} for a few lines.
A weighted mean of all measurements and their errors yields $K_2 = 4.0\,\pm\,0.8\,$\kms{}. The true error may be larger (e.g., due to pulsational variability), but is difficult to quantify.

\begin{figure*}
\centering
\includegraphics[width=\textwidth]{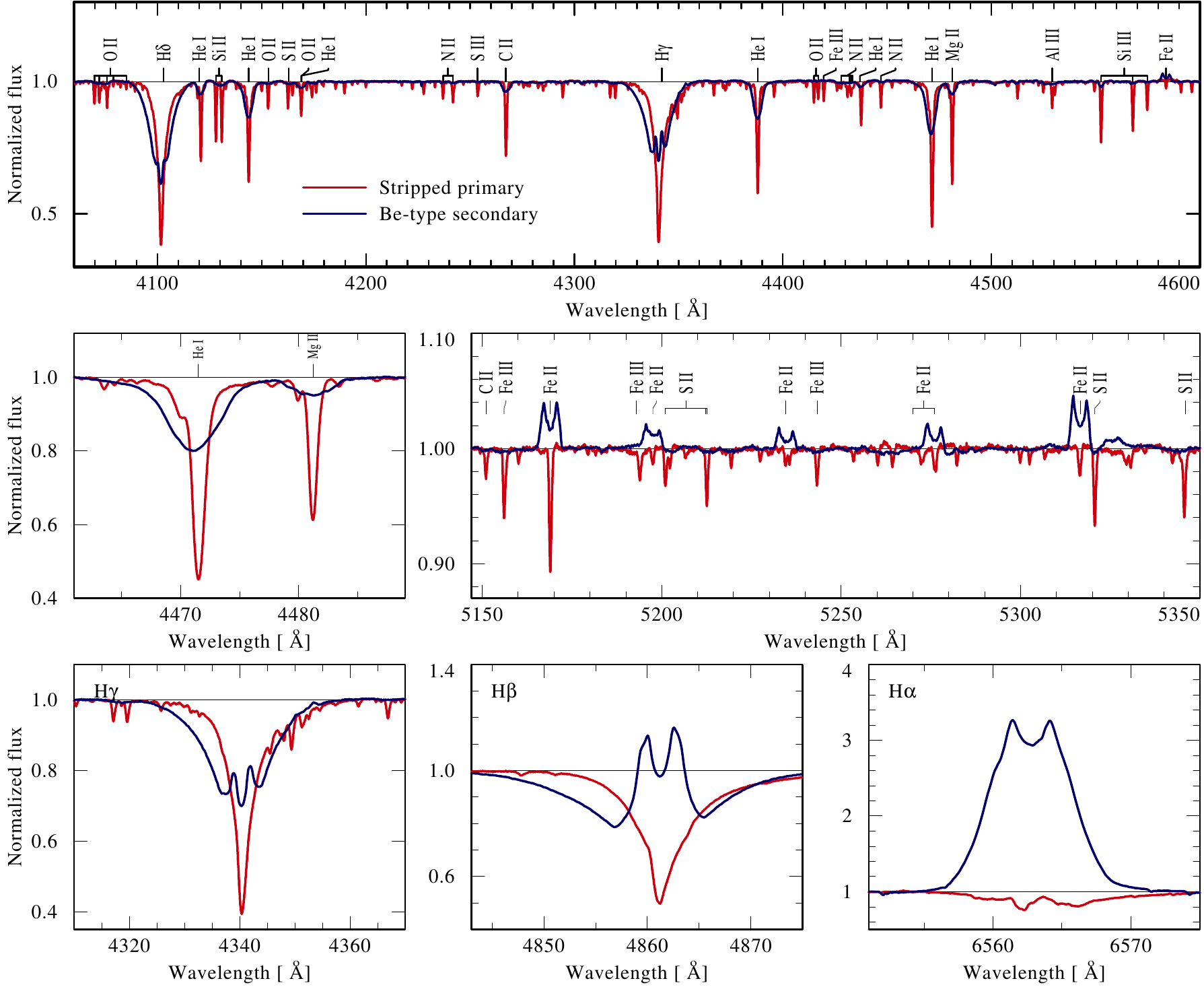}
\caption{Disentangled spectra obtained for the narrow-lined primary (red) and the rapidly rotating Be secondary (blue) using the shift-and-add algorithm.}
\label{fig:DisSpectra_ShiftandAdd}
\end{figure*}

We test these findings by applying the same overall procedure using an independent disentangling technique, i.e., Fourier disentangling \citep[][see Appendix\,\ref{sec:simulations}]{ilijicObtainingNormalisedComponent2004, Hadrava1995, ilijicFd3SpectralDisentangling2017}. Fourier disentangling of the Balmer and Fe\,\textsc{ii} emission lines results in comparable $K_2$ measurements to those derived using the shift-and-add technique (see Fig.\,\ref{fig:fd3}). The He\,{\sc i} lines favor values close to zero, but provide little diagnostic power as the $1\sigma$ confidence interval is generally larger than 7\,\kms{}. The weighted mean of the measurements and their errors of $3.7\pm1.0$\,\kms{} agrees with the shift-and-add method. A detailed comparison to the shift-and-add technique as well as an overview of the $K_2$ measurements for individual lines is given in Appendix\,\ref{sec:simulations}. 

Given the low amplitude derived we further test both techniques (shift-and-add and Fourier disentangling) using simulated spectra tailor-made for this system (see Appendix\,\ref{sec:simulations}). These tests show that in general sharp spectral features such as the double-peaked Fe\,\textsc{ii} and Balmer profiles are more sensitive to RV changes in comparison to broad, shallow rotation profiles. We find that both methods successfully retrieve the input parameters of our simulated system when disentangling the Balmer and Fe\,\textsc{ii} lines. In contrast, Fourier disentangling of the He\,\textsc{i} lines typically yields vanishing $K_2$ measurements, while the shift-and-add technique generally retrieves the input value. We note that the two disentangling procedures yield similar disentangled spectra. Hence, the spectral analysis (Sect.\,\ref{sec:specclass}) is unaffected by the choice of method.

Based on this discussion and on the results of Appendix\,\ref{sec:simulations}, we adopt $K_2 = 4.0 \pm 0.8$\,\kms{}, implying that the Be component is bound to the narrow-lined primary in the 40-d orbit.
%The question remains whether such a small value for $K_2$ can be considered as significant.
While we cannot fully rule out that variability impacts our results, the simulations using artificial data presented in Appendix\,\ref{sec:simulations} strongly suggest that our tools distinguish well between a static component and an orbital companion with a small $K_2$. At the very least, our results suggest that the data are not better explained with a static tertiary Be star, as $\chi^2(K_2=0$\,\kms{}) is significantly higher than $\chi^2(K_2=4$\,\kms{}).

% \subsection{Fourier disentangling}\label{subsec:fd3}
% Fourier disentangling relies on solving a matrix equation of the Fourier components of the composite spectra, introduced by \citet{simonDisentanglingCompositeSpectra1994, hadravaOrbitalElementsMultiple1995}. This procedure is coded into the Fourier disentangling code \texttt{fd3} (formerly \texttt{fdBinary} \citep{ilijicObtainingNormalisedComponent2004}) by \citet{ilijicFd3SpectralDisentangling2017}. Similarly to the shift-and-add technique, we fix the geometric orbital parameters $p, T_0, \omega$, e, as well as the primary semi-amplitude $K_1$, and vary the semi-amplitude $K_2$ in a grid specified by:  $K_2 \in [0, 15]$\kms, with increments of 0.1\kms. The results, using the same lines as in section \ref{subsec:shiftandadd} are presented in figure \ref{fig:GridDis_fd3}.\par
% Considering again the sharp emission features, we obtain a minimal chi-squared distance for $K_2 = 1.5 \pm 2.0$\,\kms. This is significantly lower than what the shift-and-add technique finds, and we conclude that Fourier disentangling is unsuitable for this system, as we motivate in appendix \ref{sec:simulations} with simulated data.

% \begin{figure}
% \centering
% \includegraphics[width=0.5\textwidth]{fd3_summary2.png}
% \caption{Reduced $\chi^2(K_2)$ and corresponding $K_2$ measurements for individual line groups. The red lines depict the $1\sigma$ confidence interval. }
% \label{fig:GridDis_fd3}
% \end{figure}

% \subsection{The disentangled spectra}\label{subsec:disenspec}

The disentangled spectra of the two components using $K_2=4\,$\kms{}, scaled by the estimated light ratios (see below), are shown in Fig.\,\ref{fig:DisSpectra_ShiftandAdd}. The disentangled spectra are consistent with the presence of two stellar components: a narrow-lined B-type star (primary) and a rapidly-rotating Be star (secondary), showing characteristic emission in the Balmer lines as well as in Fe\,\textsc{ii} lines. This confirms earlier studies of the system containing a narrow-lined and a broad-lined component \citep{Dachs1981, Slettebak1982, Maintz2003, Rivinius2020}.

We note that virtually all lines observed for the narrow-lined primary exhibit asymmetries, with the right wing of the lines typically being more extended than the left wing. This likely represents the collective effect of pulsational activity throughout the 2004 epoch, as suggested also in Fig.\,\ref{fig:Si_Var} in Sect.\,\ref{subsec:Si} \citep[see also][]{Aerts2009}. Moreover, the H$\alpha$ line of the primary, and to a lesser extend the H$\beta$ line, shows indications for emission in its core. The lack of significant absorption in H$\alpha$ is consistent with the lack of significant variability observed in the core of H$\alpha$ in the individual observations (see Sect.\,\ref{subsec:halpha_variability}). While the observed emission in H$\alpha$ could be intrinsic to the primary (e.g., due to stellar winds), it could also represent an interaction between the primary and the Be disk (e.g., disk irradiation) that follows the orbit of the primary.

The light ratio cannot be determined from the disentangling process. We therefore estimate it from the two disentangled spectra.  We find that both spectra can be reasonably reproduced with models when assuming a light contribution of $\approx$\,45\% for the primary and $\approx$55\,\% for the secondary. The light contributions cannot differ from these values by more than $\approx 20\%$, because the scaled disentangled spectra would otherwise become unphysical\footnote{ For example, assuming a light contribution less than $30\%$ for the primary would cause its Balmer absorption lines in the scaled spectrum to reach negative flux values.}. We thus fix this light ratio in the further analysis, and note that small changes in the light ratio adopted do not have a strong impact on the determination of the stellar parameters and our final conclusions.

\section{Spectral Analysis}\label{sec:specclass}
\subsection{The primary narrow-lined B star}\label{subsec:Bprim}
\subsubsection{Comparison to the B3III giant 18\,Peg}
\begin{figure} \centering
\includegraphics[width=0.5\textwidth]{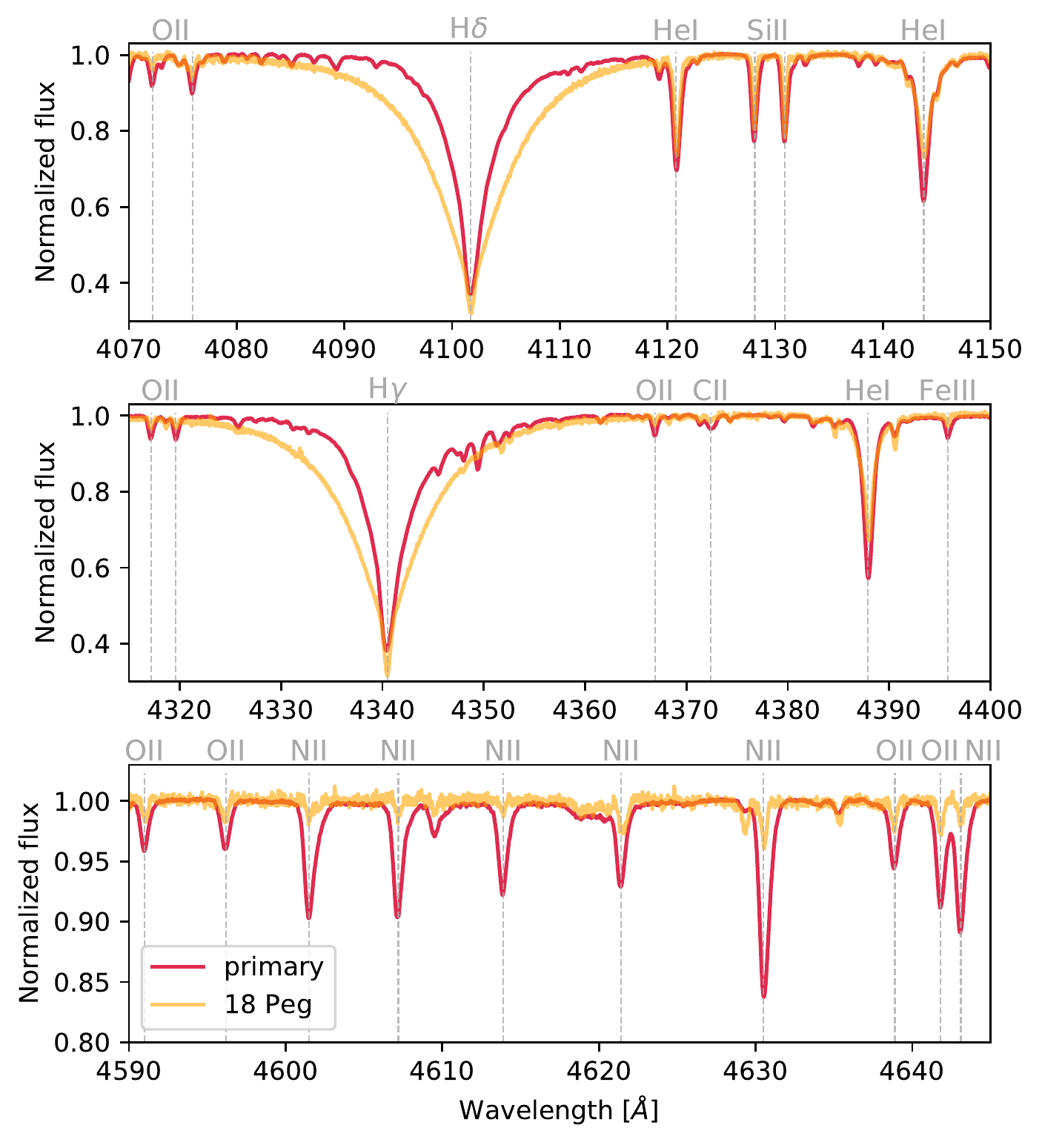}
\caption{Comparison of the disentangled primary spectrum scaled for its flux contribution (red) to the spectrum of the prototypical B3 giant 18\,Peg (orange). Both spectra are shifted to rest wavelength.}
\label{fig:18Peg}
\end{figure}

\cite{Rivinius2020} reported that the B-type primary is a fairly normal B3III giant and hence adopt stellar parameters derived for the prototypical B3III giant, 18\,Peg \citep{Nieva2014}. We thus qualitatively compare the disentangled and scaled spectrum of the primary to the spectrum of 18\,Peg observed with the {\sc HERMES} spectrograph \citep{Raskin2011} mounted on the Mercator telescope (wavelength coverage: 3770 to 9000\,$\AA$, spectral resolving power: $\sim$\,85\,000) in Fig.\,\ref{fig:18Peg}.

The comparison shows that, while the spectra are in general similar, the Balmer wings of \Starname{} are much narrower than the ones of 18\,Peg. This indicates a lower $\log g$ in \Starname{}. Furthermore it can be seen that O\,\textsc{ii} and especially N\,\textsc{ii} lines are significantly deeper in \Starname{}. However, without a detailed spectral analysis, it is not readily clear whether this is due to anomalous surface abundances in the primary of \Starname{} or due to different values of $T_\mathrm{eff}$ and $\log g$.

\subsubsection{Rotation and macroturbulence}\label{subsubsec:linebroad}
The line profiles of the narrow-lined primary are dominated by a "triangular" shape from a radial-tangential broadening profile \citep{Gray1973, Simon-Diaz2014}. This implies that the dominant broadening mechanism is not rotation, but macroturbulent velocity (%which has been concluded to be
caused by non-radial pulsations, see \citealt{Aerts2009, Bowman2020}; and Sect.\,\ref{subsec:Si}). For slow rotators, and especially in cases where the projected rotational velocity \vsini{} and the macroturbulent velocity \vmac{} are comparable, the two parameters are highly degenerate \citep[see e.g.,][]{Simon-Diaz2007}. 

\begin{figure} \centering
\includegraphics[width=.5\textwidth]{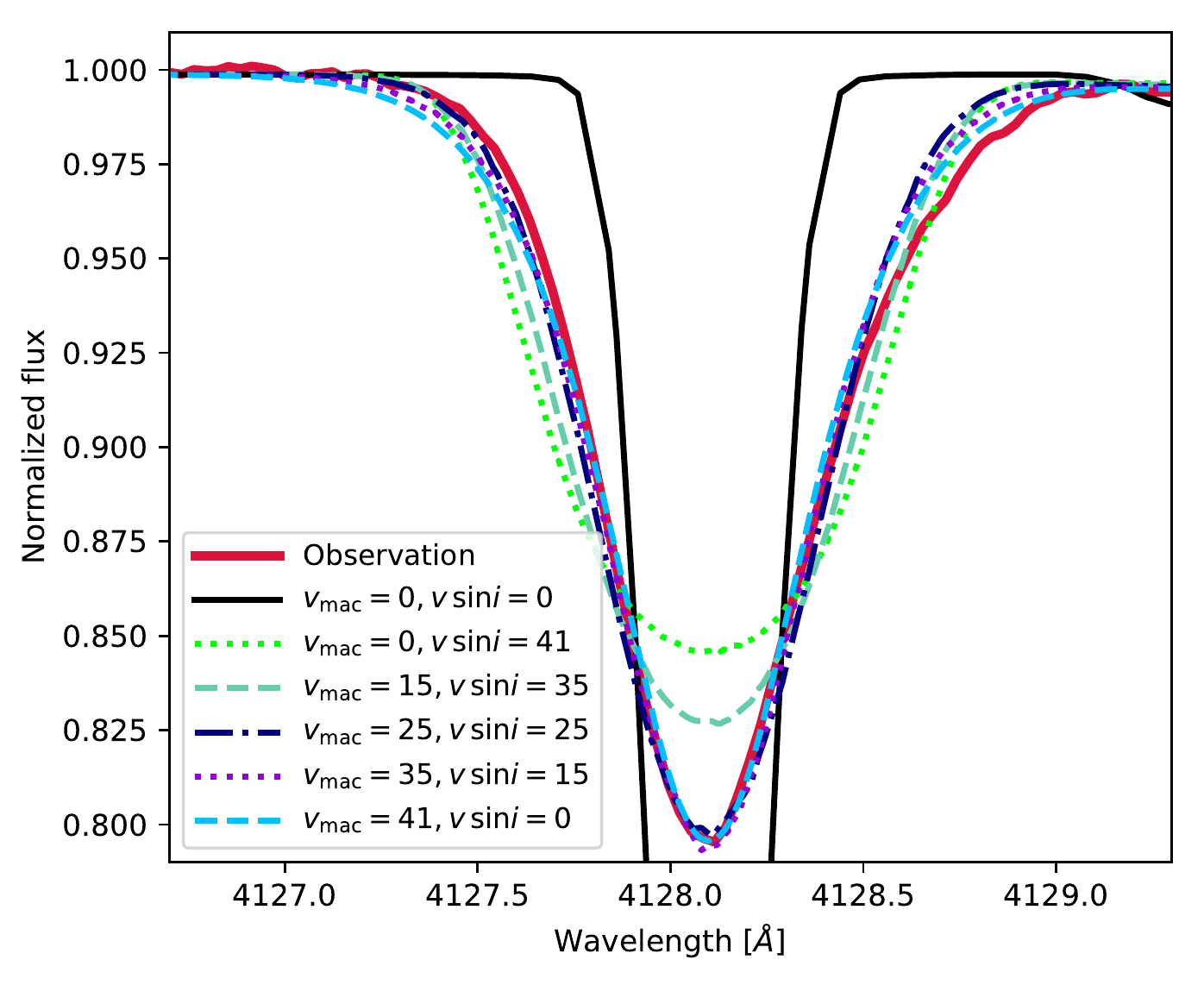}
\caption{Comparison between the disentangled spectrum of the Si\,{\sc ii}\,$\lambda 4128$ line with a PoWR model atmosphere for the narrow-lined primary calculated with \vmic{}$= 10\,$\kms{} but with different rotation and macroturbulent velocities. The values are given in the legend in \kms{}.}
\label{fig:vmacvsini}
\end{figure}

To estimate the broadening parameters we therefore perform a visual comparison between the disentangled spectrum and a model atmosphere calculated with the Potsdam Wolf-Rayet (PoWR) code (see Sect.\,\ref{subsec:teff1}). Figure\,\ref{fig:vmacvsini} illustrates such a comparison for the Si\,{\sc ii}\,$\lambda 4128$ line. It can be seen that, within uncertainties related to both the disentangling procedure and the modeling of the macroturbulence, one cannot distinguish between the fit quality of a non-rotating and a slowly rotating star. Figure\,\ref{fig:vmacvsini} also illustrates that \vsini{} values larger than $\approx\,25$\kms{} can be excluded. We adopt \vsini{}=15\,\kms{} and \vmac{}=35\,\kms{} for the subsequent analysis, but we note that the true \vsini{} value may be lower owing to the spectral resolving power.

% We test the derived parameters with the {\sc iacob-broad} tool \citep{Simon-Diaz2007, Simon-Diaz2014} on the N\,{\sc ii}\,$\lambda 3995$, N\,{\sc ii}\,$\lambda 4601$, Si\,{\sc ii}\,$\lambda 4128$, and Si\,{\sc iii}\,$\lambda 4553$ lines in the disentangled spectrum. 
% The tool relies on the Fourier method \cite{Gray1973} and goodness-of-fit for determining \vsini{} and \vmac{}. The tool reports typical values of \vsini{} $\approx 17\,$\,\kms{} and \vmac{} $\approx 40\,$\,\kms{}, which agree well with the values derived above. 
% However, applying this tool on simulated spectra of a non-rotating star with \vmac{} $= 40\,$\kms{} and \vmic{} $=10\,$\kms{} yield comparably high \vsini{} values. From this we conclude that \vsini{} and \vmac{} are highly degenerate \cite[see also discussion in][]{Simon-Diaz2007}, and we refrain from using \textsc{iacob-broad} here.

\subsubsection{Stellar parameters and abundances}\label{subsec:teff1}
We compare the disentangled and scaled spectrum of the primary to the BSTAR2006 grid of
the model atmosphere code \textsc{tlusty} \citep{Hubeny1995, Lanz2007}, which  solves the radiative transfer problem in plane-parallel geometry without assuming local thermodynamic equilibrium (non-LTE). The grid covers effective temperatures between 15 and 30\,kK in steps of 1\,kK and surface gravities between 1.75 and 4.75 in steps of 0.25 dex. The microturbulent velocity is fixed to \vmic{}=2.0\,\kms{}. We find that the Balmer, helium and most metal lines are well reproduced with models corresponding to $T_\mathrm{eff}\approx$ 16-17\,kK and $\log g \approx 2.8$\,\cgs. However, several metal lines, especially both Si\,\textsc{ii} and Si\,\textsc{iii} lines, are strongly underestimated by the models. Assuming that the abundance of heavy elements is solar, this indicates that the \vmic{} value is higher than 2\,\kms.

We therefore use an appropriate grid of atmosphere models calculated with the non-LTE Potsdam Wolf-Rayet (PoWR) code \citep{Graefener2002, Hamann2003, Sander2015}. The grid used here is based on an extension of a B-star grid calculated for solar metallicity by \citet{Hainich2019}, but with negligible mass-loss rates ($\log \dot{M} \lesssim -9.0\,[M_\odot\,{\rm yr}^{-1}]$). It covers the relevant parameter regime: $15 \le T_{\rm eff} \le 19\,$kK  with a spacing of $\Delta T_{\rm eff} = 1\,$kK, $2.4\le \log g \le 3.2\,$\,\cgs{} with a spacing of 0.2\,dex, and microturbulent velocities of 2, 5 and 10\,\kms{}. 

We find that both \vmic{}= 5\,\kms{} and \vmic{}= 10\,\kms{} reproduce the spectra well, with that latter value yielding better results. We therefore fix \vmic{}= 10\,\kms{} in the further analysis, noting that the true value may be anywhere between $\approx5$ to $\approx15\,$\kms{}. As possible abundance changes significantly alter the depth of spectral lines, we avoid using temperature diagnostics based on two different elements such as the ratio between He\,\textsc{I} and Mg\,\textsc{II} lines. Instead, we use the ratio between Si\,\textsc{ii} to Si\,\textsc{iii} as prime temperature diagnostic. As the secondary has Fe\,\textsc{ii} lines in emission (see Fig.\,\ref{fig:DisSpectra_ShiftandAdd}), which might propagate as additional uncertainty in the line profile of the primary, we refrain from using Fe\,\textsc{ii} to Fe\,\textsc{iii} ratios. By comparing the measured ratios of equivalent widths (EWs) of the observed spectrum to those measured in PoWR models, we find possible T$_\mathrm{eff}$-$\log g$ combinations that reproduce the observed Si line ratios within their errors. 
%In Fig.\,\ref{fig:logg} we compare these to the primary spectrum.

% As -solar abundances of certain elements (see below) that significantly change the depth of the lines, we refrain from using temperature diagnostics such as the ratio between He\,\textsc{I} and Mg\,\textsc{II} lines. We rather use different ionization stages of the same element to remove a possible impact of altered abundances.

\begin{figure}
\centering
\includegraphics[width=0.5\textwidth]{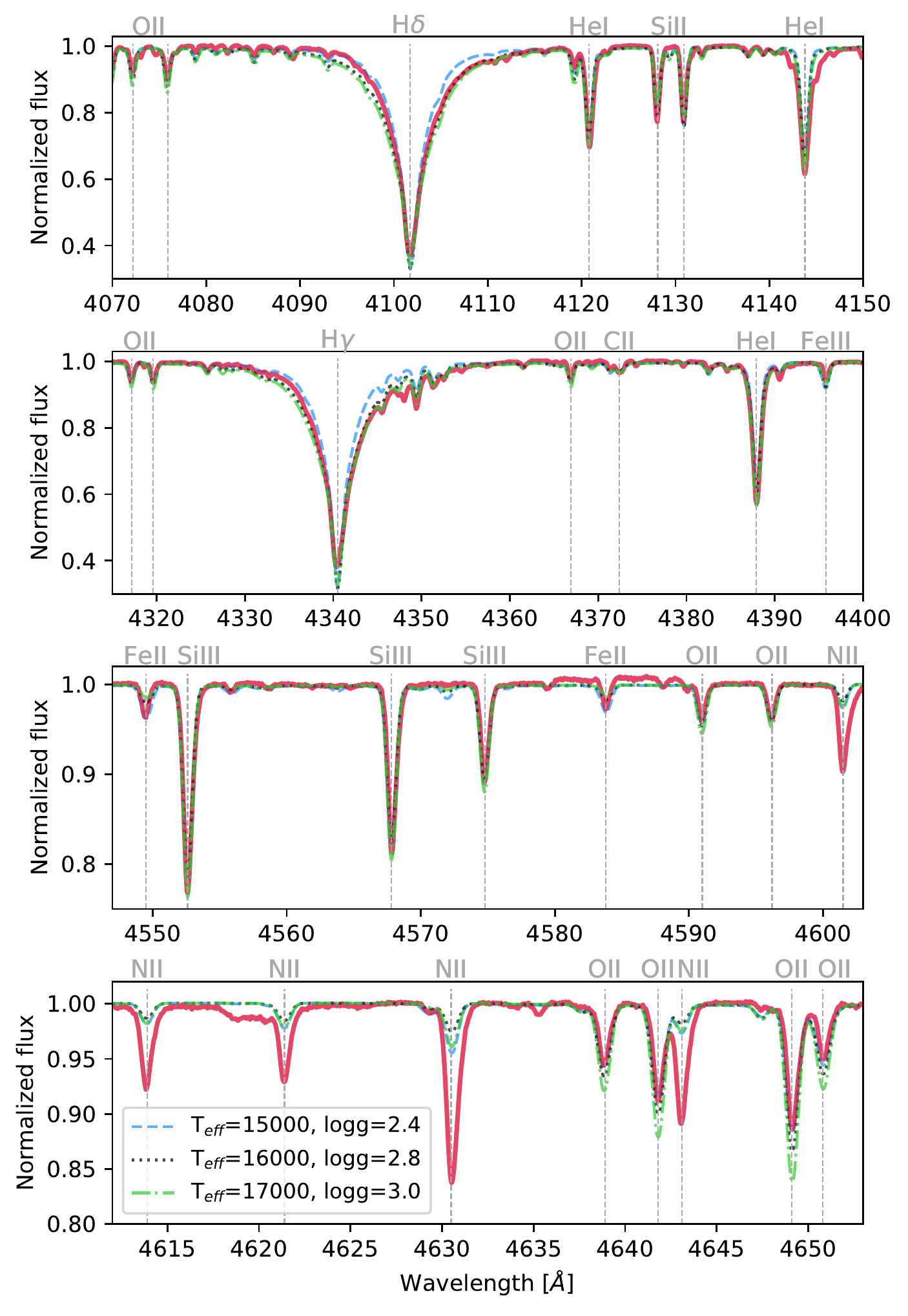}
\caption{Comparison between the disentangled primary spectrum scaled for its flux contribution (red) with models for possible T$_\mathrm{eff}$-$\log g$ combinations (see legend for the respective T$_\mathrm{eff}$ and $\log g$ values).}
\label{fig:logg}
\end{figure}

Figure\,\ref{fig:logg} shows a comparison between three best-fitting PoWR models with the disentangled spectrum of the primary scaled for its flux contribution. It illustrates that the overall primary spectrum is well reproduced by the model spectra. This is also true for the wings of Balmer and He\,\textsc{i} lines, which are sensitive to the disentangling procedure and might thus be subject of uncertainty. Given the good overall agreement we argue that this uncertainty is rather small.

% While our temperature estimate agrees with the estimate by \cite{Rivinius2020}, the $\log g$ we find here is significantly lower than the $\log g$ the authors assume to be typical of a B3III giant (i.e. $\log g$ $\approx$ 3.75 \citep{?}).
% %This indicates that the primary might not be a star towards the end of its main-sequence evolution.

The bottom panel of Fig.\,\ref{fig:logg} focuses on several N and O lines. It shows that the N\,\textsc{ii} lines are consistently deeper in the observations than in the models. The opposite effect occurs for the O\,\textsc{ii} lines, which are weaker in the observations. While this is a qualitative comparison and an in-depth abundance analysis is required, it is an indication for an abundance pattern in line with expectations for CNO-processed material. %The He\,\textsc{i} lines seem slightly stronger in the observations.
There is, however, no strong indication for an elevated He abundance. We notice the same trend in the abundance pattern of the CNO elements when using \textsc{tlusty}.

% We test our findings with two additional atmosphere codes: \textsc{tlusty} \citep{Hubeny1995, Lanz2007} and \textsc{fastwind} \citep{Santolaya1997, Puls2005, Rivero2011}. The BSTAR2006 grid of \textsc{tlusty} models is only available with \vmic{}=2.0\,\kms{}. A comparison of those with the observed spectrum indicates again that the star has a higher microturbulent velocity. The obtained effective temperature and surface gravity agree within the uncertainties. By comparing \textsc{fastwind} and PoWR models, some differences are identified. However, we find again find similar results for T$_\mathrm{eff}$, $\log g$ and \vmic{} values. Furthermore, we notice the same trend in the abundance pattern of the CNO elements irrespective of the code used. 

% \begin{figure} \centering
% \includegraphics[width=0.5\textwidth]{CNO.pdf}
% \caption{Comparison between the disentangled spectrum of the primary (red) with PoWR models for possible T$_\mathrm{eff}$-$\log g$ combinations (see legend for the respective T$_\mathrm{eff}$ and $\log g$ values).}
% \label{fig:abundances}
% \end{figure}

\subsection{The secondary Be component}\label{subsec:Besec}

\subsubsection{Comparison to the spectra of classical Be stars}
\begin{figure} 
\centering
\includegraphics[width=0.5\textwidth]{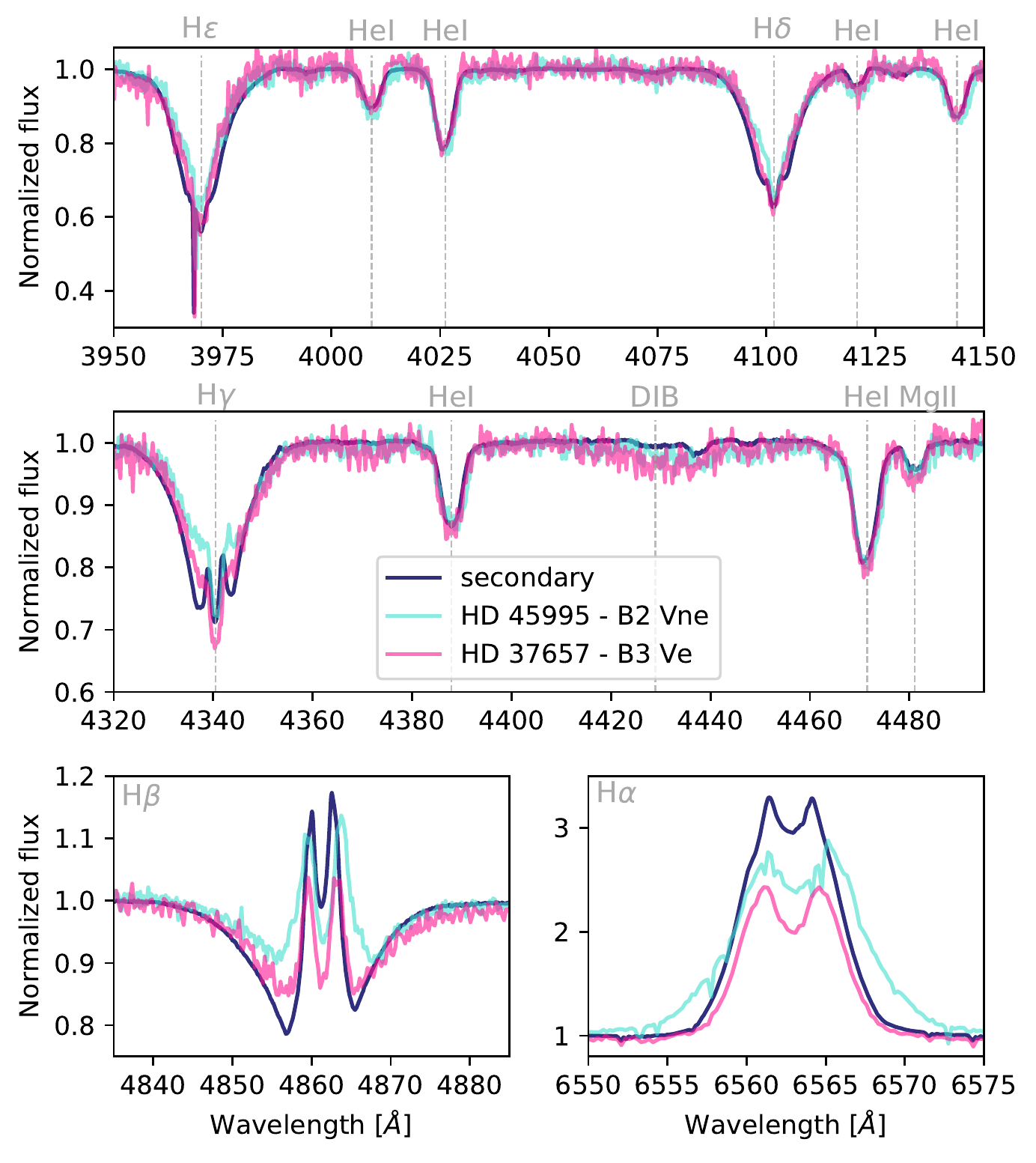}
\caption{Comparison between the disentangled spectrum of the secondary (blue) with other representative Be type stars observed with \textsc{HERMES}. The spectra are binned at $\Delta \lambda=0.2\,\AA$ for clarity.}
\label{fig:secondary_Bes}
\end{figure}

Similarly to the primary, we first qualitatively compare the secondary's spectrum (scaled to the 55\% light contribution) to spectra of representative classical Be stars, again observed with \textsc{HERMES}. In Fig.\,\ref{fig:secondary_Bes} we show the comparison to HD\,45995 \citep[B2Vne,][]{Jaschek1982} and HD\,37657 \citep[B3\,Ve,][]{Guetter1968}, for which \cite{Zorec2016} report \vsini{} values of $255\pm20$\,\kms{} and $198\pm17$\,\kms{}, respectively. In general, the spectrum of the secondary is well reproduced by the Be star spectra. A comparison of the Mg\,\textsc{ii} to the He\,\textsc{i} line indicates that the secondary is probably of spectral type B2-3~V. 

The morphology of the Balmer emission, which primarily depends on the properties of the circumstellar disk and its inclination, is better matched the spectrum of HD\,37657 (B3\,Ve). The morphology of the \halpha{} line is in agreement with a relatively low inclination \citep[$i \lesssim 45^\circ$, e.g.,][]{Rivinius2013}, and is potentially affected by disk truncation induced by the companion \citep{Negueruela2001, Okazaki2002}.

\subsubsection{Rotation and macroturbulence}\label{subsubsec:linebroad2}

We determine the projected rotational velocity \vsini{} and the macroturbulent velocity \vmac{} of the Be star using the {\sc iacob-broad} tool \citep{Simon-Diaz2007, Simon-Diaz2014}. As the He\,{\sc i} lines are significantly affected by pressure broadening and strong variability, we use the relatively isolated and strong metal lines C\,{\sc ii}\,$\lambda 4267$ and Mg\,{\sc ii}\,$\lambda 4482$. For the C\,{\sc ii} line, the Fourier method yields \vsini = $183\,$\kms{}, while the goodness-of-fit yields \vsini{} $= 184\pm10$\,\kms{} and \vmac $= 70\pm30$\,\kms{}. For the Mg\,{\sc ii} line, the Fourier method yields \vsini = $166\,$\kms{}, while the goodness-of-fit yields \vsini{} $= 167\pm20$\,\kms{} and \vmac = $65\pm50$\,\kms{}. Both lines therefore yield comparable results. A weighted average yields \vsini = $180\pm10$\,\kms~and \vmac = $70\pm25$\,\kms.
We note that both lines contain a blend of two spectral lines with separations of $\approx 0.2\,\AA$, such that the derived broadening values may be slightly overestimated. 

The high \vmac{} value is consistent with the strong variability observed in the He\,{\sc i} lines (Sect.\,\ref{subsec:HeI}). Assuming the rotational axis of the Be star is aligned with the orbital inclination of $i=32^\circ$ (see Sect.\,\ref{subsec:lum_mass}), the \vsini{} value derived here implies an equatorial rotation of $\varv_{\rm eq}\approx 340\,$\kms{} for the Be component, which amounts to roughly $80$\% of its estimated critical velocity \citep{Townsend2004}.
 
\subsubsection{Stellar parameters}
We compare the disentangled spectrum of the secondary to a grid of \textsc{TLUSTY} models with \vmic{}= 2\,\kms{}, which allows us to reproduce the spectrum well. We broaden the spectra according to the \vsini{} and \vmac{} described above.

The Balmer and some metal lines such as Fe\,\textsc{ii} are affected by emission and their depth can not be used as diagnostics. We thus resort to the He\,\textsc{i} lines and the ratio of the He\,{\sc i}\,$\lambda 4472$ to the Mg\,{\sc ii}\,$\lambda 4481$ line as a temperature diagnostic. For $\log g$ we focus on the wings of the Balmer and He\,{\sc i} lines.

\begin{figure} \centering
\includegraphics[width=0.5\textwidth]{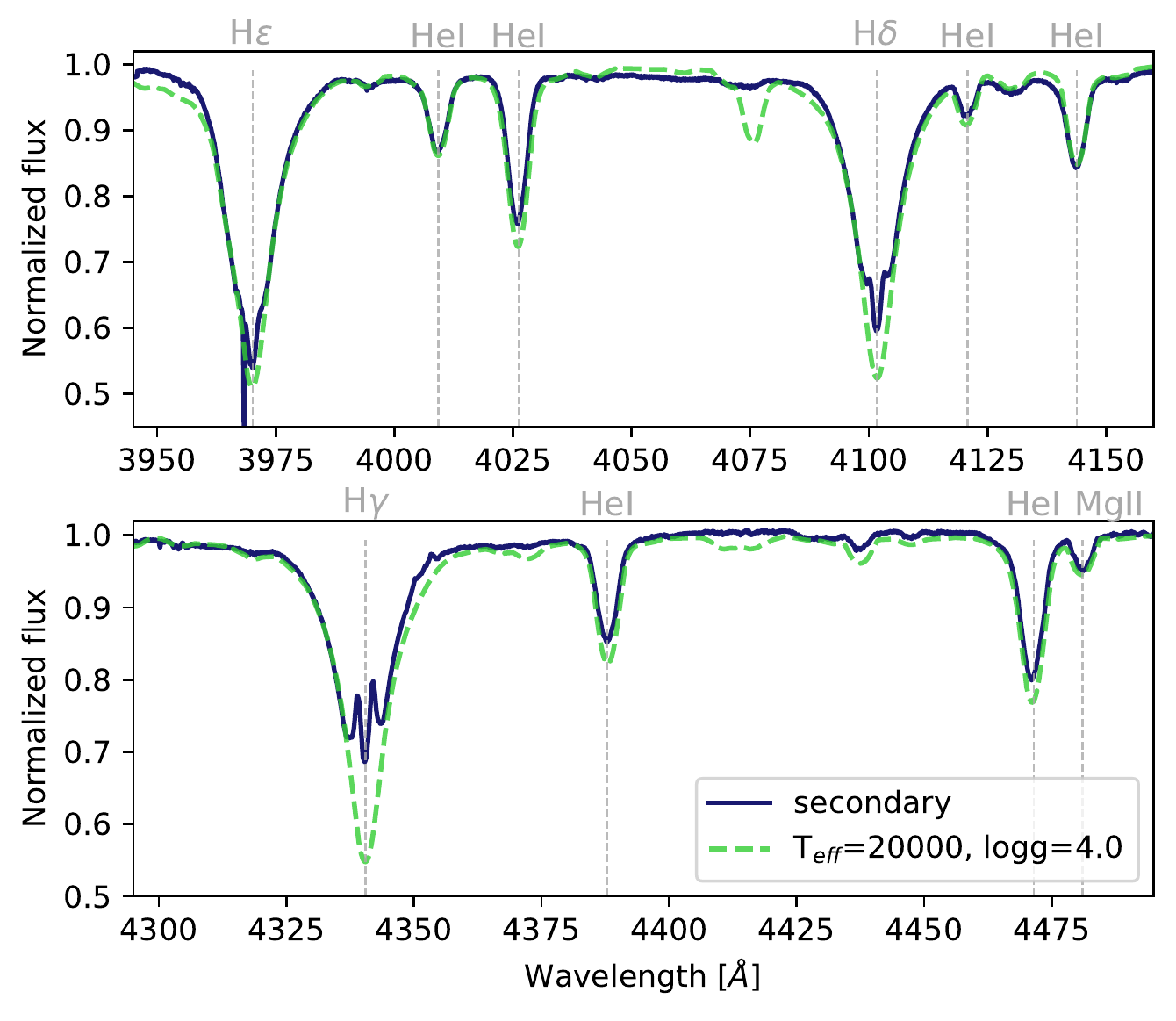}
\caption{Comparison between the disentangled spectrum of the secondary (blue) with the best-fit TLUSTY model (green).}
\label{fig:secondary_spec}
\end{figure}

We find that the secondary is best reproduced by a model with $T_{\rm eff} \simeq 20\,000\,$K and a $\log g$ of $\sim$4.0\,\cgs{} (see Fig.\,\ref{fig:secondary_spec}). These parameters agree well with the spectral type B2-3~V derived above.

\section{Doppler Tomography}\label{sec:tomography}
The spectra of \Starname{} contain an \halpha{} emission line profile reminiscent of that found in Be star spectra. \citet{Rivinius2020} proposed that this emission is attributed to a static Be star that is located further away from the 40-d binary system. Based on our measurement of $K_2$, we argue here that the Be star is in orbit with the primary on a 40-d period.

To test the consistency of our scenario with the data, we further investigate the origin and distribution of the \halpha{} emissivity by applying the Doppler tomography technique to the \halpha{} line profile. It maps the formation region of these lines in velocity space and requires that the observations fully cover the orbital cycle. We therefore only apply this technique to the 2004 data since the data from 1999 were only collected during half an orbit. 	

As described in \citet{Mahy2012}, the Doppler tomography technique assumes that the radial velocity of any gas flow is stationary in the rotating frame of reference of the system. The velocity of a gas parcel, as seen by the observer, can be expressed by a so-called `S-wave' function, 
\begin{eqnarray}
v(\phi) = -v_x \cos(2\pi \phi) + v_y \sin(2\pi \phi) + v_z
\end{eqnarray}
where $\phi$ represents the orbital phase, ($v_x,v_y$) are the velocity coordinates of the gas flow and $v_z$ is the systemic velocity. This relation assumes an $x-$axis situated between the stars, from the primary to the secondary, and a $y-$axis pointing in the same direction as the orbital motion of the secondary star \citep[see e.g.,][]{rauw02,linder08,Mahy2012}. To reconstruct the images, our method uses a Fourier-filtered back-projection algorithm. 

\begin{figure} \centering
\includegraphics[width=0.5\textwidth]{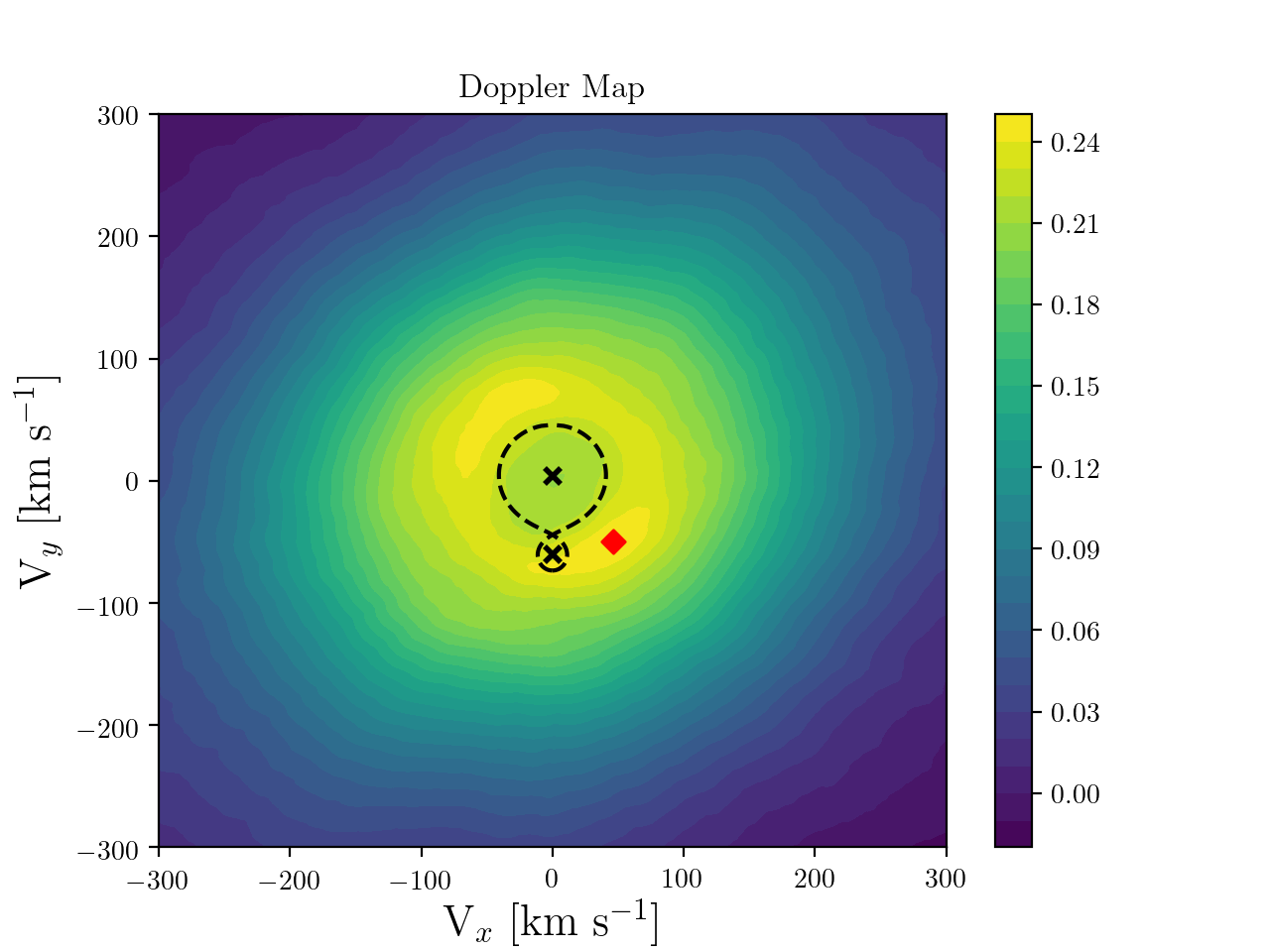}
\caption{Doppler map of the \halpha{} line of \Starname{}. The crosses correspond to the radial velocity amplitudes of the primary and secondary stars. The shape of the Roche lobe in velocity space (thick dashed line) was calculated for a mass ratio (secondary/primary) of 15.1. The red diamond indicates the maximum of emissivity.}
\label{fig:doppler}
\end{figure}

The Doppler map (Fig.\,\ref{fig:doppler}) was computed assuming the radial velocity semi-amplitudes listed in Table\,\ref{tab:Bstar_properties} ($K_1 = 60.4$\,\kms{}, $K_2=4.0$\,\kms{} and $\gamma = 9.13$\,\kms{}). We have also represented in this map the Roche lobes for the primary (narrow-lined star) and the secondary (Be star). The ring-shaped nature of the emissivity is a signature of an accretion or decretion disk located around the secondary star of the 40-d period system. This is consistent with the secondary being a Be star or a compact object with an accretion disk. In the latter case, a significant X-ray luminosity is expected which was not observed by \citet{Berghoefer1996}, and thus excluded by \citet{Rivinius2020}.

While the Doppler map indicates the viability of the B+Be model on a 40-d orbit, it does not allow us to test the scenario where the Be star is a distant third companion as the phase coverage provided by the existing data is insufficient to compute a reliable Doppler map in such a scenario. %To summarise, the Doppler map provides no insight of the actual position of the Be star in the system.}

\section{Evolutionary discussion}\label{sec:discussion}
\subsection{Spectroscopic and dynamical masses}\label{subsec:lum_mass}

The spectroscopic masses can be calculated from $M_{\rm spec}\,\propto\,g\,R^2$, where the stellar radius $R$ can be inferred from the Stefan-Boltzmann equation $R \propto L^{1/2}\,T_\mathrm{eff}^{-2}$, with $L$ the stellar luminosity. With $\log g$ and $T_{\rm eff}$ already inferred for both components,  
we estimate their luminosities by comparing the observed spectral energy distribution (SED) of \Starname{} with the sum of two PoWR models calculated with the same parameters as those given in Table\,\ref{tab:Bstar_properties}, fixing the light ratio to 45\% and 55\% in the visual for the narrow-lined primary and Be secondary, respectively (see Sect.\,\ref{subsec:Bprim}). We use the reddening law given by \citet{Cardelli1989} with a total-to-selective extinction ratio of $R_V = 3.1$. For the distance, we adopt the Gaia measurement of $d = 340\pm20$\,pc provided by \citet{Bailer-Jones2018}. The relatively low astrometric excess noise (0.7\,mas), the high parallax-over-error value (15), and the re-normalised unit weight error (RUWE) lower than 1.4 \citep[1.14,][]{Lindegren2018}, suggest that, despite being a binary, the measurement of the parallax is not immediately problematic and that the distance estimate is trustworthy.
Finally, we adopt UBVI photometric measurements from \citet{Ducati2002} and JHK photometry from \citet{Cutri2003}. Moreover, we use two available flux calibrated UV spectra\footnote{IDs: SWP31220, SWP31220, PI: T.\ Snow} obtained with the International Ultraviolet Explorer (IUE). 

\begin{figure}
\centering
\includegraphics[width=.5\textwidth]{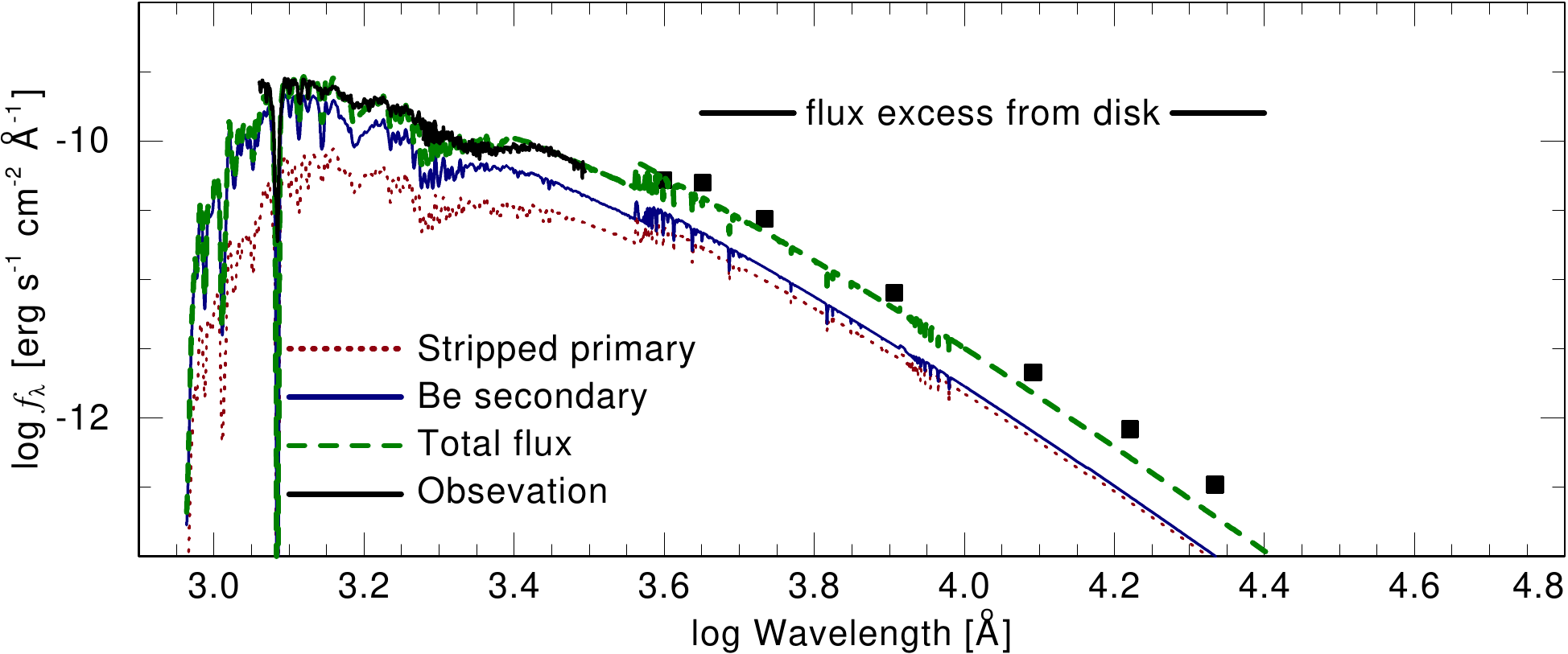}
\caption{Comparison between observed photometry and IUE spectra of \Starname~and the model SED calculated as the sum of two PoWR models with parameters corresponding to those given in Table\,\ref{tab:Bstar_properties} (see legend). }
\label{fig:SED}
\end{figure}

The SED comparison is shown in Fig.\,\ref{fig:SED}. We find a good agreement for a reddening of $E_{B-V} = 0.1$\,mag and luminosities of $\log L/L_\odot = 3.05$ and $3.35$ for the primary and secondary, respectively. Significant flux excess is observed in the infrared and to a lesser extent in the visual, which most likely stems from the disk of the Be component \citep[e.g.,][]{Cochetti2019}.  

Using the derived luminosities, we estimate the spectroscopic masses. Interestingly, the spectroscopic mass of the narrow-lined component turns out to be $M_{\rm spec, 1} = 0.4^{+0.3}_{-0.1}$\,\Msun{}. This value is clearly not in agreement with the primary being a standard B-type main sequence (MS) star, as assumed by \citet{Rivinius2020}. Instead, the mass, together with the enhanced N-abundance, suggests that the narrow-lined B-type component is another example of a thermally-unstable stripped star, similarly to \object{LB-1} \citep{Irrgang2020, Shenar2020}. Moreover, combining this mass estimate with the binary mass function $f_{\rm M} = 0.96$\,\Msun{} implies that the minimum mass of the companion on the 40-d orbit is $\approx$1.5\,\Msun{}. Hence, even if the secondary is not the Be star, as proposed here, but rather an unseen companion, one cannot immediately rule out the possibility that it is a faint MS star or a neutron star, instead of a BH.

The luminosity of the Be star is broadly consistent with its estimated spectral type of B2-3\,V \citep{Hohle2010}. The spectroscopic mass of the secondary is found to be around 15 times larger than the one of the primary, i.e., $M_{\rm spec, 2} =  6_{-3}^{+5}$\,\Msun{}, where the conservative $\log g$ error is used. This mass is also in agreement with typical values reported for Be stars of comparable spectral classes \citep{Hohle2010}. 

By calibrating the mass of the secondary Be star to a mass typical for its spectral type, $M_2 = 7\pm2$\,\Msun{} \citep{Hohle2010}, we can obtain a measurement of the orbital inclination and the dynamical mass of the narrow-lined star, which is independent of the spectroscopic mass estimated above. This yields an inclination of $i=32\pm4^{\circ}$ and $M_\mathrm{1,dyn} = 0.46\pm0.24$\,\Msun{}, in excellent agreement with our spectroscopic mass estimate. 

In summary, the present analysis supports a formation scenario for \Starname{} similar to that proposed for {LB-1} by \citet{Shenar2020}, involving a past binary mass-exchange event that led to the formation of the stripped star and its rapidly-rotating Be companion.

\subsection{Evolutionary history}\label{subsec:somethingsomethingevolution}
If the system is indeed composed of a bloated stripped star that recently donated mass to the Be star, its current orbital properties can be used to assess the initial conditions of the progenitor system. Assuming circular orbits, a constant mass transfer efficiency, and ignoring other mechanisms for angular momentum loss from the system, the orbital period can be computed analytically as a function of the mass ratio \citep{Soberman+1997}. For the case of conservative mass transfer one has that
\begin{eqnarray}
\frac{P}{P_r}=\left(\frac{q}{q_r}\right)^{-3}\left(\frac{1+q}{1+q_r}\right)^6, \label{eq:cons_mt}
\end{eqnarray}
where $P_r$ and $q_r\equiv M_{\rm Be}/M_{\rm stripped}$ represent the orbital period and mass ratio at a particular reference point, which we take to be the currently observed values. In the case where mass transfer is fully non-conservative, with the ejected mass carrying away the specific angular momentum of the accreting star, the evolution of the orbital period is given by\footnote{The definition of $q$ in Eq.\,3 is the inverse of the one used in \citet{Soberman+1997}. It also corrects a typo in Eq. 22 of that work.}
%Note that this corrects a typo in equation B22 of \citet{Soberman+1997} and that our definition of $q$ is also the inverse of the one used in that work.}
\begin{eqnarray}
\frac{P}{P_r}=\left(\frac{q}{q_r}\right)^{5}\left(\frac{1+q}{1+q_r}\right)^{-2}e^{3(q_r-q)/q_r q}.\label{eq:nc_mt}
\end{eqnarray}
Both of these expressions allow us to determine the orbital period of the system at a previous point in time, when its mass ratio was smaller.

\begin{figure}
\centering
\includegraphics[width=.5\textwidth]{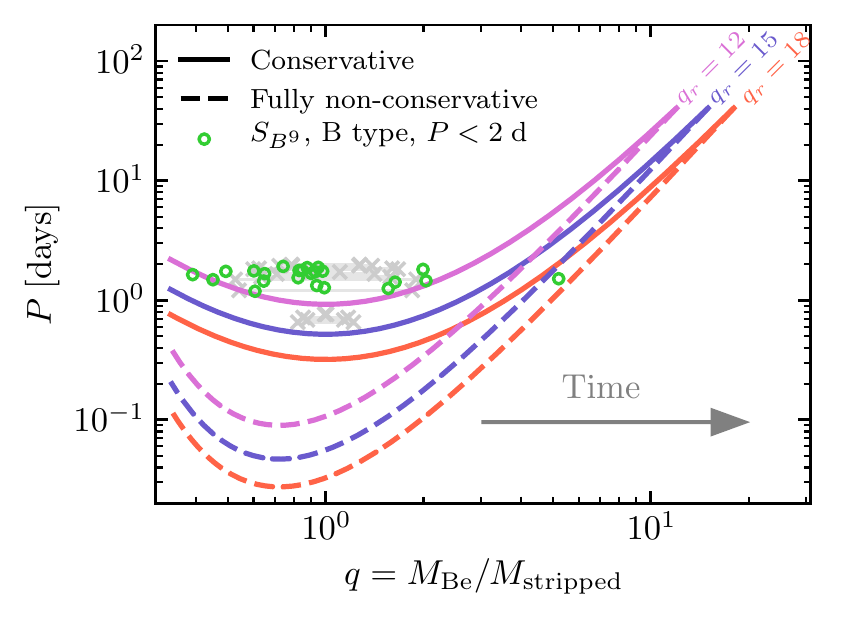}
\caption{Past orbital evolution of the system assuming present-day mass ratios of $q=12,15$ and $18$. During mass transfer phases time increases towards the right in this Figure. For each curve the rightmost point represents the currently observed values, while the left-most point corresponds to an arbitrary choice of $q=1/3$. Circles are a selection of detached and semidetached systems from the $S_{B^9}$ catalogue. Crosses are systems from the same catalogue for which it is unclear if the binary has undergone a mass ratio inversion, so two points are included per system with inverse values for $q$ (see text for details).}
\label{fig:mass_ratio_evol}
\end{figure}
Figure\,\ref{fig:mass_ratio_evol} shows the result of evaluating Eqs.\,(\ref{eq:cons_mt}) and (\ref{eq:nc_mt}) for different values of the currently observed mass ratio within our range of uncertainty. Although we do not know the mass ratio at birth, the stripped star had to be the most massive component ($q<1$ with our definition of mass ratio). If we assume the initial mass ratio was $q=1/3$, then the initial orbital period needs to be $\lesssim$\,2~d to explain its current value. Moreover, fully non-conservative mass transfer requires initial orbital periods smaller than 10~hr. For either case, the short initial periods required are indicative of the system having undergone Roche-lobe overflow during the main sequence (Case A mass transfer). This can help explain the extreme mass ratio of the system, as Case A evolution leads to smaller masses for the resulting stripped stars compared to binaries that interact after the main sequence \citep{Wellstein+2001}. 

To check for the existence of known binary systems that can be potential progenitors for \Starname{}, we consider all double-lined binaries contained in the $S_{B^9}$ catalogue of spectroscopic binaries \citep{Pourbaix2004} with a period of less than 2~d. In addition, to exclude systems with either too high or too low total masses, we only take into account binaries with one B-type component, and without an O-type component. Although this gives us enough information to determine the mass ratios, we do not know if the system has interacted and potentially undergone a mass ratio inversion. To remove this uncertainty we consider detached binaries that would correspond to a pre-interaction system with $q<1$, and semi-detached binaries with a Roche-lobe filling secondary that correspond to the phase after mass inversion with $q>1$. For all other binaries that do not fall into either of these categories we cannot assess if there has been a mass ratio inversion, so we consider both possibilities for the mass ratio (see Appendix \ref{sec:sb9list} for a full list of the systems). As shown in Fig.\,\ref{fig:mass_ratio_evol}, there are various observed systems that, through conservative mass transfer, can lead to a binary with a mass ratio comparable to the one we derive for \Starname{}. Notably, there are no detached systems that could serve as progenitors through non-conservative mass transfer. 

\begin{figure}
\centering
\includegraphics[width=.5\textwidth]{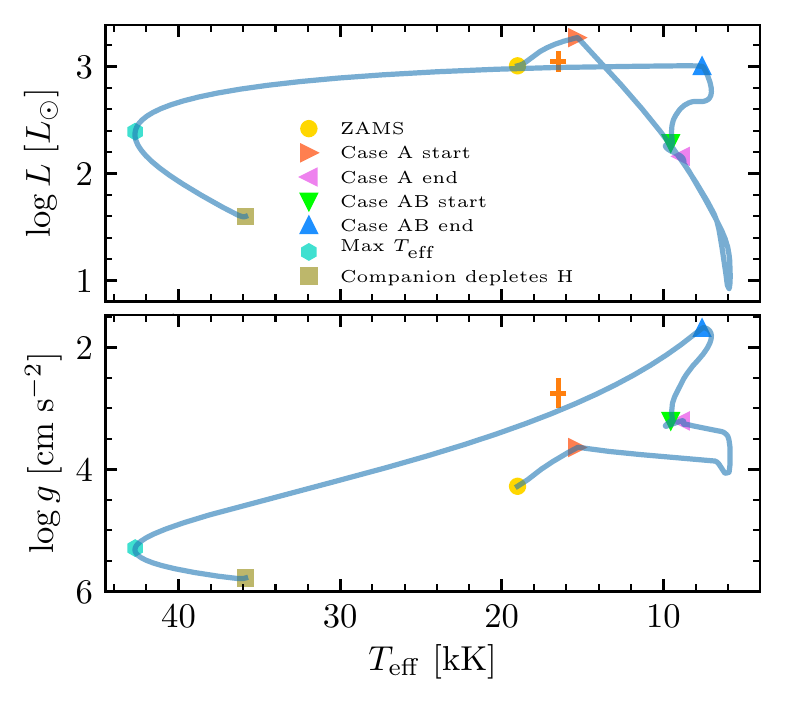}
\caption{Evolution in the Hertzsprung-Russel (top panel) and Kiel diagrams (bottom panel) of the donor star of a binary system with initial masses of $6\,M_\odot$ and $2\,M_\odot$ and orbital period of $2$ days. The orange error bars indicate the observed values of the B star in \Starname{}, and different symbols indicate different evolutionary stages (see legend).}
\label{fig:bin_evol}
\end{figure}

Using Eqs. (\ref{eq:cons_mt}) or (\ref{eq:nc_mt}) as a guide we can choose initial properties of the system to attempt to model it using a stellar evolution code. As a simple example, we consider here a binary system with initial masses of $6\,M_\odot$ and $2\,M_\odot$ at a period of 2~d undergoing fully conservative mass transfer. The model is computed using the \texttt{MESA} code for stellar evolution \citep{Paxton+2011,Paxton+2013,Paxton+2015,Paxton+2018,Paxton+2019}. Full details of our simulation settings and physics assumptions, as well as necessary steps to reproduce this simulation, are provided in Appendix \ref{sec:mesa}. The evolution of the donor star in this system is illustrated in Fig. \ref{fig:bin_evol}. The star evolves unperturbed until it fills its Roche-lobe after 55 Myrs of evolution. At this point the star is still on the main sequence, resulting in a phase of Case A mass transfer that lasts until the donor depletes its central hydrogen. This phase lasts 23 Myrs and reduces the mass of the donor to $1.5\,M_\odot$. After core-hydrogen depletion, another phase of mass transfer ensues (commonly referred to as Case AB mass transfer) that lasts for 6.7 Myrs and reduces its mass to $0.59\,M_\odot$. The star then detaches and evolves towards increasing $T_{\rm eff}$, reaching its maximum value after $1$ Myr. For the last 12 Myrs of the simulation the stripped star burns helium at its core, but before it depletes it, the accreting star, which now has a mass of $7.4\,M_\odot$, finishes its main sequence evolution. The further evolution of the model is not computed, but a phase of inverse mass transfer is expected which, owing to the extreme mass ratio, would be unstable.

In this simulation, the evolutionary stage that corresponds to the B star in \Starname{} is right after detachment from Case AB mass transfer. When the stripped star in the model reaches $T_{\rm eff}=16$ kK, it has $\log L/L_\odot=2.99$ and $\log g=2.97$\,\cgs, while the companion has $T_{\rm eff}=22$ kK, $\log L/L_\odot=3.57$ and $\log g=4.34$\,\cgs. The mass ratio at this stage is $q=12.5$ and the orbital period is of $38.9$ days. All these values are comparable to the currently observed ones, despite not having done an extensive parameter search to find a progenitor system. One important aspect that appears to be inconsistent with the observations though, is that the mass fraction of helium at the surface of the stripped star is predicted to be $Y=0.87$, whereas no helium enhancement is observed in spectroscopy. It is unclear which physical processes could explain this discrepancy. Gravitational settling can lower the abundance of elements heavier than hydrogen at the surface \citep{IbenMacdonald1985}, but including this effect in our simulation shows that it does not operate on the brief $1$ Myr phase of contraction. A more speculative scenario is that mass being lost by the Be star through a decretion disk contaminates the surface of the stripped star with material that is hydrogen-rich.

Given the similarities of our model to the observed system, one can use it to roughly estimate the rarity of observing such systems with a bloated stripped star right after mass transfer. By simply taking ratios of the lifetimes of the different phases, compared to the $1$ Myr period after mass transfer when the star evolves to higher $T_{\rm eff}$, one finds that for each system such as \Starname{}, among the progenitor systems one expects 55 pre-interaction binaries, 23 systems undergoing mass transfer on the MS, about 7 systems undergoing case AB mass transfer, and 11 systems composed of a compact core-helium burning star and a rapidly rotating B star. The latter are potentially observable as a single Be star given the low mass and low luminosity of the core-helium burning companion. The post-mass transfer phase that corresponds to \Starname{} represents about 1\% of the total lifetime, so for every system similar to \Starname{} one would expect about $100$ systems in one of the other evolutionary stages. While rare, this is by no means an exceptional occurrence rate. In the future, we expect the \Starname{} system to merge when the Be star depletes the hydrogen in its core, owing to the extreme mass ratio of the system.

\section{Conclusions} \label{sec:concl}
We performed a detailed spectroscopic and evolutionary analysis of \Starname, which was recently proposed by \citet{Rivinius2020} to be a hierarchical triple system comprising a BH + B-giant binary on a 40-d orbit and a distant Be tertiary.
Based on the available spectroscopic observations of \Starname{}, our spectral disentangling implies that the spectral features of the Be star (Balmer lines, photospheric He\,{\sc i} lines, and disk Fe\,{\sc ii} and O\,{\sc ii} lines) move in anti-phase with the narrow spectral features of the B-type primary. We derive RV amplitudes of $K_1 = 60.4\pm1.0\,$\kms~and $K_2=4.0\pm0.8\,$\kms, which translates to an extreme mass ratio between the two components of $M_2 / M_1 \approx 15\pm3$.
%The Doppler tomography performed gives further evidence that the disk in which the \halpha{} emission originates is associated with the secondary of the 40-d orbit and not with a component on a wider orbit. 
We thus conclude that the Be star is not a tertiary, but instead the secondary in the 40-d period system.  

Our spectroscopic analysis shows that the primary has T$_\mathrm{eff}\approx16\pm1$\,kK and $\log\,g\approx 2.8\pm0.2$\,\cgs. We find a high macroturbulent velocity of \vmac{}= $35\pm5$\,\kms{} and can only report an upper limit on the rotational velocity \vsini{}$\lesssim25$\,\kms{}. The secondary spectrum resembles the typical spectrum of a Be star and the comparison to classical Be stars indicates that it is of spectral type B2-3~Ve. This agrees with the estimated atmospheric parameters of T$_\mathrm{eff}\approx20\pm2$\,kK and $\log\,g\approx 4.0\pm0.3$\,\cgs. We derive \vsini{} = 185$\pm20\,$\kms{} and \vmac{} = 70$\pm 25$\,\kms{}. The comparably high values of \vmac{} in both stars are in line with the observation of strong variability in both components on different timescales, with observed short- and medium-term variability consistent with non-radial pulsations \citep{Aerts2009, Rivinius2020}.

Based on luminosities derived from a SED fit, we estimate the spectroscopic masses of the two stars. We find that the spectroscopic mass of the primary is $M_{\rm spec, 1} = 0.4^{+0.3}_{-0.1}\,M_\odot$ while the spectroscopic mass of the secondary is $M_{\rm spec, 2} =  6_{-3}^{+5}\,M_\odot$. %This, together with the enhanced nitrogen abundance, implies that the narrow-lined primary is a thermally-unstable stripped star.
Our analysis suggests that \Starname{} is a binary system containing a stripped star and a Be star, bearing similarity to the recently discovered binary \object{LB-1} \citep{Liu2019, Irrgang2020, Shenar2020}. By modelling the system with \textsc{MESA}, we show that we can reproduce the currently observed parameters with a tight progenitor system that has undergone Case A mass transfer. One important caveat of our model is that we find no indications for the strong enrichment in helium which is predicted by the models. We do, however, observe indications for an enhanced nitrogen abundance which agrees with the predictions.

This work shows that the \Starname{} system can be explained without invoking the presence of a stellar-mass BH. Considering the low value of $K_2$ derived here and the time coverage of the data,  we cannot fully exclude that the Be star is stationary (Appendix\,\ref{sec:simulations}), nor can we fully exclude the presence of additional components on an even shorter orbit (Sect.\,\ref{subsec:HeI}). A final test to reveal the configuration of the system and the nature of its components can only be given by an independent measurement of the separation between the visible components. Given the small distance to the system, a combination of interferometric and high-contrast imaging observations could help to test all different scenarios. This opens an unprecedented observing window in the important parameter space to constrain the nature of this intriguing system.

\begin{acknowledgements}
Based on observations obtained with the HERMES spectrograph, which is supported by the Research Foundation - Flanders (FWO), Belgium, the Research Council of KU Leuven, Belgium, the Fonds National de la Recherche Scientifique (F.R.S.-FNRS), Belgium, the Royal Observatory of Belgium, the Observatoire de Genève, Switzerland and the Thüringer Landessternwarte Tautenburg, Germany. This work has made use of the BeSS database, operated at LESIA, Observatoire de Meudon, France: http://basebe.obspm.fr. We thank T. Bohlsen and P. Luckas for uploading their spectra to the BeSS database. The research leading to these results has received funding from the European Research Council (ERC) under the European Union's Horizon 2020 research and innovation programme (grant agreement numbers 772225: MULTIPLES and 670519: MAMSIE), the FWO Odysseus program under project G0F8H6N and the FWO junior fellowship 12ZY520N. PM would like to thank Alina Istrate for helpful discussion.
\end{acknowledgements}

\bibliographystyle{aa}
\bibliography{papers}

\appendix

\section{Journal of the observations}
Table \ref{tab:obs_log} provides a journal of observations including exposure times, signal-to-noise (S/N) ratios and RVs of the narrow-lined primary. In Table \ref{tab:obs_log_bess} we give an overview over the additional BeSS spectra used together with their exposure times and resolving powers.
\begin{table}[h!]\centering
\caption{\label{tab:obs_log} Journal of the observations of \Starname{}. The second column indicates the exposure time while the third column gives the S/N ratio in the spectrum at $\approx 5400\,\AA$. The last column gives the RV of the primary star.}
    \begin{tabular}{llrr} \hline \hline 
     $\mathrm{HJD}$\,[d] & time [s] & S/N & RV$_1$ [\kms{}]\\ \hline
    % \begin{longtable}{llrr}
    % \hline 
    %  $\mathrm{HJD}$\,[d] & time [s] & S/N & RV$_1$ [\kms{}]\\
    %  \hline
    %  \endfirsthead
    %  \caption{\it continued.} \\
    %  \hline \hline 
    %  $\mathrm{HJD}$\,[d] & time [s] & S/N & RV$_1$ [\kms{}]\\
    %  \hline
    %  \endhead
    %  \hline
    %  \endfoot
2451373.676 & 300 & 150 & $-$39.0 $\pm$ 4.5 \\
2451374.675 & 300 & 290 & $-$43.6 $\pm$ 4.5 \\
2451375.653 & 300 & 300 & $-$41.6 $\pm$ 4.5 \\
2451376.724 & 1800 & 200 & $-$30.9 $\pm$ 4.5 \\
2451378.682 & 1500 & 280 & $-$27.7 $\pm$ 4.4 \\
2451380.824 & 360 & 240 & $-$14.8 $\pm$ 4.4 \\
2451383.673 & 300 & 270 & 12.1 $\pm$ 4.4 \\
2451384.743 & 300 & 230 & 24.1 $\pm$ 4.4 \\
2451385.759 & 300 & 200 & 32.8 $\pm$ 4.5 \\
2451390.745 & 300 & 300 & 62.9 $\pm$ 4.9 \\
2451393.627 & 300 & 200 & 68.9 $\pm$ 5.1 \\
2451394.758 & 300 & 220 & 65.5 $\pm$ 5.0 \\ \hline
2453138.847 & 150 & 320 & $-$7.7 $\pm$ 4.4 \\
2453139.682 & 150 & 300 & $-$14.5 $\pm$ 4.4 \\
2453143.808 & 150 & 230 & $-$50.0 $\pm$ 4.6 \\
2453144.753 & 150 & 340 & $-$54.4 $\pm$ 4.6 \\
2453149.762 & 150 & 320 & $-$51.1 $\pm$ 4.6 \\
2453149.765 & 150 & 320 & $-$51.1 $\pm$ 4.6 \\
2453154.669 & 150 & 300 & $-$21.5 $\pm$ 4.4 \\
2453159.804 & 150 & 340 & 18.5 $\pm$ 4.4 \\
2453160.686 & 150 & 370 & 28.5 $\pm$ 4.5 \\
2453162.862 & 150 & 230 & 42.0 $\pm$ 4.5 \\
2453185.551 & 150 & 400 & $-$52.6 $\pm$ 4.6 \\
2453187.666 & 150 & 330 & $-$51.6 $\pm$ 4.6 \\
2453188.687 & 150 & 360 & $-$53.7 $\pm$ 4.6 \\
2453190.704 & 150 & 310 & $-$49.3 $\pm$ 4.6 \\
2453194.691 & 150 & 350 & $-$25.6 $\pm$ 4.4 \\
2453195.563 & 150 & 350 & $-$14.7 $\pm$ 4.4 \\
2453196.550 & 150 & 270 & $-$2.7 $\pm$ 4.4 \\
2453197.670 & 150 & 330 & 8.1 $\pm$ 4.4 \\
2453199.570 & 150 & 310 & 26.8 $\pm$ 4.5 \\
2453202.645 & 150 & 330 & 51.4 $\pm$ 4.7 \\
2453204.513 & 150 & 340 & 66.3 $\pm$ 5.0 \\
2453205.697 & 150 & 320 & 69.8 $\pm$ 5.2 \\
2453207.501 & 150 & 250 & 74.9 $\pm$ 5.3 \\
2453226.596 & 150 & 290 & $-$52.0 $\pm$ 4.6 \\
2453230.490 & 150 & 420 & $-$45.9 $\pm$ 4.5 \\
2453231.569 & 150 & 360 & $-$40.5 $\pm$ 4.5 \\
2453232.502 & 150 & 300 & $-$39.7 $\pm$ 4.5 \\
2453239.494 & 150 & 420 & 21.2 $\pm$ 4.4 \\
2453240.602 & 150 & 300 & 29.6 $\pm$ 4.5 \\
2453243.515 & 150 & 400 & 56.9 $\pm$ 4.8 \\
2453244.551 & 150 & 370 & 62.4 $\pm$ 4.9 \\
2453245.510 & 150 & 430 & 67.0 $\pm$ 5.1 \\
2453246.619 & 150 & 310 & 64.7 $\pm$ 5.0 \\
2453247.553 & 150 & 320 & 64.4 $\pm$ 5.0 \\
2453248.522 & 150 & 430 & 69.4 $\pm$ 5.1 \\
2453254.539 & 150 & 300 & 48.0 $\pm$ 4.6 \\
2453255.515 & 150 & 290 & 36.5 $\pm$ 4.5 \\
2453256.555 & 150 & 250 & 27.7 $\pm$ 4.5 \\
2453257.534 & 150 & 350 & 21.0 $\pm$ 4.4 \\
2453258.492 & 150 & 350 & 16.3 $\pm$ 4.4 \\
2453259.531 & 150 & 400 & 3.6 $\pm$ 4.4 \\
2453260.536 & 150 & 380 & $-$5.3 $\pm$ 4.4 \\
2453261.541 & 150 & 370 & $-$13.7 $\pm$ 4.4 \\
2453262.530 & 150 & 400 & $-$21.3 $\pm$ 4.4 \\
2453263.536 & 150 & 420 & $-$35.7 $\pm$ 4.5 \\
2453264.567 & 150 & 340 & $-$42.6 $\pm$ 4.5 \\ \hline
    \end{tabular}
\end{table}

\begin{table}[h!]\centering
\setcounter{table}{0}
    \caption{\it continued.}
    \begin{tabular}{llrr} \hline \hline 

     $\mathrm{HJD}$\,[d] & time [s] & S/N & RV$_1$ [\kms{}]\\ \hline
2453264.567 & 150 & 340 & $-$42.6 $\pm$ 4.5 \\
2453265.490 & 150 & 370 & $-$44.8 $\pm$ 4.5 \\
2453269.508 & 150 & 310 & $-$52.2 $\pm$ 4.6 \\
2453271.514 & 150 & 310 & $-$44.7 $\pm$ 4.5 \\
2453272.540 & 150 & 410 & $-$37.5 $\pm$ 4.5 \\
2453273.505 & 150 & 360 & $-$28.7 $\pm$ 4.4 \\ \hline
    \end{tabular}
\end{table}

\begin{table}[h!]\centering
\setcounter{table}{1}
    \caption{Journal of the observations of \Starname{} from the BeSS. The second column indicates the exposure time while the third column gives approximate resolving power.}
    \label{tab:obs_log_bess}
    \begin{tabular}{lll} \hline \hline 
     $\mathrm{HJD}$\,[d] & time [s] & resolving power \\ \hline
2455821.045 & 3400 & 3000 \\
2456168.065 &  300 & 600 \\
2456923.978 &  100 & 1000 \\
2457274.066 & 2400 & 15000 \\
2457963.044 & 1200 & 15000 \\
2458335.035 & 1200 & 15000 \\
2458775.957 & 1200 & 15000 \\
2458981.000 & 1800 & 15000 \\ \hline

    \end{tabular}
\end{table}

\section{1999 vs.\ 2004}\label{sec:1999}
We attempt to derive an orbital solution using the 1999 data alone. We assume $e=0$ and the same period given in Table\,\ref{tab:Bstar_properties},  and vary the remaining orbital parameters. Doing so, we find that the 1999-data yield slightly different results. We find % a non-zero eccentricity ($e=0.05 \pm 0.03$) and
$K_1 = 54.2 \pm 1.2$\,\kms{} (instead of 60.4 $\pm$ 1.0\,\kms{}). This is probably due to the low number of spectra in 1999 (12 epochs) that cover only half the orbital period.

\begin{figure}
\centering
\includegraphics[width=0.5\textwidth]{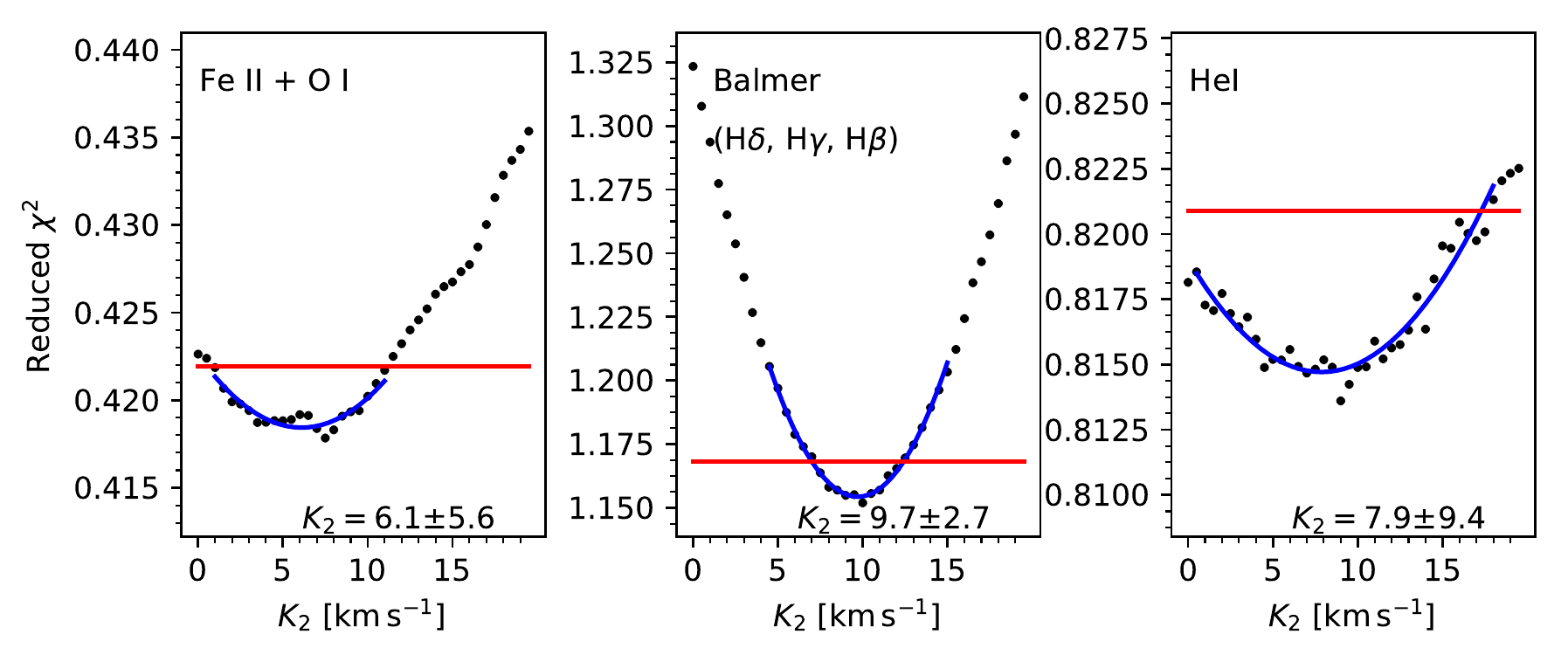}
\caption{Like Fig.\,\ref{fig:MasterGridDis_Main}, but using the  12 spectra of the 1999 epoch instead of the 51 spectra of the 2004 epoch.}
\label{fig:1999Dis}
\end{figure}

Similarly, we try to perform disentangling along the $K_2$-axis using the 12 observations taken during the 1999 epoch alone (see Sect.\,\ref{sec:disen}). The results for the He\,{\sc i} lines, Fe\,{\sc i} + O\,{\sc ii} lines, and Balmer lines (excluding \halpha{} for consistency with the 2004 epoch) are shown in Fig.\,\ref{fig:1999Dis}. Evidently, using the 1999 epoch alone, larger values of $K_2$ are preferred. The Fe and He measurements lie slightly above those of the 2004 measurements (cf.\,Fig.\,\ref{fig:MasterGridDis_Main}), but well within their respective 1$\sigma$ errors. The Balmer lines imply a significantly larger $K_2$ in the 1999 epoch, but the corresponding $K_2$ values are consistent within $2\sigma$. A weighted average of all measurements for the 1999 epoch yields $K_2 = 8.9\pm2.4\,$\kms{}. This is significantly larger than the 2004 measurement of $K_2 = 4.0\pm0.8\,$\kms{}, but consistent within 2$\sigma$.

A value of $K_2 \approx 9\,$\kms~would not change the core of our conclusions, i.e., that \Starname{} is a binary comprising a stripped primary and a rapidly rotating Be secondary. It would, however, imply a less extreme mass ratio of $M_2/M_1 \approx 7$ and a more massive B-type primary with $M_1 \approx 1\,M_\odot$.  
The short time coverage of the observations and the very strong variability observed during the 1999 epoch may well impact the validity of the results obtained for the 1999 epoch. 
Regardless, the analysis of the 1999 epoch further illustrates that evidence for a static Be star, as proposed by \citet{Rivinius2020}, is lacking.

% moved figures here to put them on the same page

\section{Are the $K_2$ measurements significant?}\label{sec:simulations}

\setlength{\tabcolsep}{6pt}
\renewcommand{\arraystretch}{1.2}
\begin{table}
\centering 
\caption{$K_2$ measurements derived from shift-and-add and Fourier disentangling of individual lines, along with the corresponding values of the reduced $\chi^2$ obtained at minima.}
    \begin{tabular}{lrr} \hline \hline 
    line & shift-and-add & Fourier \\
     &   \multicolumn{2}{c}{$K_2$ [\kms]} \\ \hline
    %  line & $K_2$ [\kms]  & $K_2$ [\kms]  \\ \hline
% H$\alpha$ & $1.9\pm1.2$ & 39.9 \\
H$\beta$ & $3.5\pm1.3$  & $4.5^{+1.5}_{-2.0}$ \\
H$\gamma$ & $5.3\pm2.2$  & $5.5^{+3.5}_{-3.0}$ \\
H$\delta$ & $3.3_{-3.3}^{+4.6}$ &  $7.5^{+5.0}_{-7.5}$ \\ \hline
Fe\,{\sc ii}\,$\lambda 4584$ & $4.1\pm4.0$ & $3.0^{+4.0}_{-3.0}$ \\
Fe\,{\sc ii}\,$\lambda 5169$ & $3.5\pm3.0$ & $4.0\pm4.0$  \\ 
Fe\,{\sc ii}\,$\lambda 5198$ & $3.2_{-3.2}^{+5.9}$ & $3.5^{+7.5}_{-3.5}$ \\ 
Fe\,{\sc ii}\,$\lambda 5235$ & $8.4\pm5.0$ & $6.0^{+8.0}_{-6.0}$  \\ 
Fe\,{\sc ii}\,$\lambda 5276$ & $3.8\pm3.8$  & $7.0^{+5.0}_{-7.0}$ \\ 
Fe\,{\sc ii}\,$\lambda 5317$ & $3.8\pm2.2$ & $3.0\pm3.0$ \\ 
Fe\,{\sc ii}\,$\lambda 5363$ & $0.5_{-0.5}^{+11.1}$ & $9.0^{+7.0}_{-9.0}$ \\ 
O\,{\sc i}\,$\lambda 8446$ & $4.2\pm2.2$ & $1.5^{+3.5}_{-1.5}$ \\ \hline
He\,{\sc i}\,$\lambda 4009$ & $0.0^{ + 19.1}$ & $0.0^{ + 10.0 }$\\ 
He\,{\sc i}\,$\lambda 4026$ & $5.3_{-5.3}^{+9.2}$  & $0.0^{+ 9.5}$ \\ 
He\,{\sc i}\,$\lambda 4122$ & $13.2_{-13.2}^{+13.8}$  & $0.0^{+ 3.5}$ \\ 
He\,{\sc i}\,$\lambda 4144$ & $2.5_{-2.5}^{+6.9}$ & $0.5^{+12.0}_{-0.5}$ \\ 
He\,{\sc i}\,$\lambda 4388$ & $4.6_{-4.6}^{+8.7}$  & $1.0^{+19.0}_{-1.0}$ \\ 
He\,{\sc i}\,$\lambda 4472$ & $2.6_{-2.6}^{+11.9}$  & $0.0^{+ 7.5}$ \\ 
\hline
weighted mean & $4.0\pm0.8$ & $3.7 \pm 1.0$    \\
\hline
    \end{tabular}
\label{tab:K2Meas}    
\end{table}

%We perform a similar disentangling procedure as described in Sect.\,\ref{sec:disen}  based on the Fourier technique. Here, we use \texttt{fd3} \citep[formerly \texttt{fdbinary},][]{ilijicObtainingNormalisedComponent2004, ilijicFd3SpectralDisentangling2017} which operates in Fourier space. The results of the Fourier disentangling are given in Fig.\,\ref{fig:fd3} which shows the reduced $\chi^2(K_2)$ and corresponding $K_2$ measurements for the individual line groups. It is important to note that the reduced $\chi^2$ is computed in Fourier space and is thus not directly comparable with the one computed with the shift-and-add technique.

Table\,\ref{tab:K2Meas} gives an overview over the $K_2$ measurements of individual lines obtained by shift-and-add and the Fourier method. It shows that for Fe\,\textsc{ii} and Balmer lines, both techniques yield similar results. The measurements based on He\,\textsc{i} lines, however, favor values close to zero in the Fourier technique which is in contrast to the measurements of the shift-and-add technique.

\begin{figure}
\centering
\includegraphics[width=0.5\textwidth]{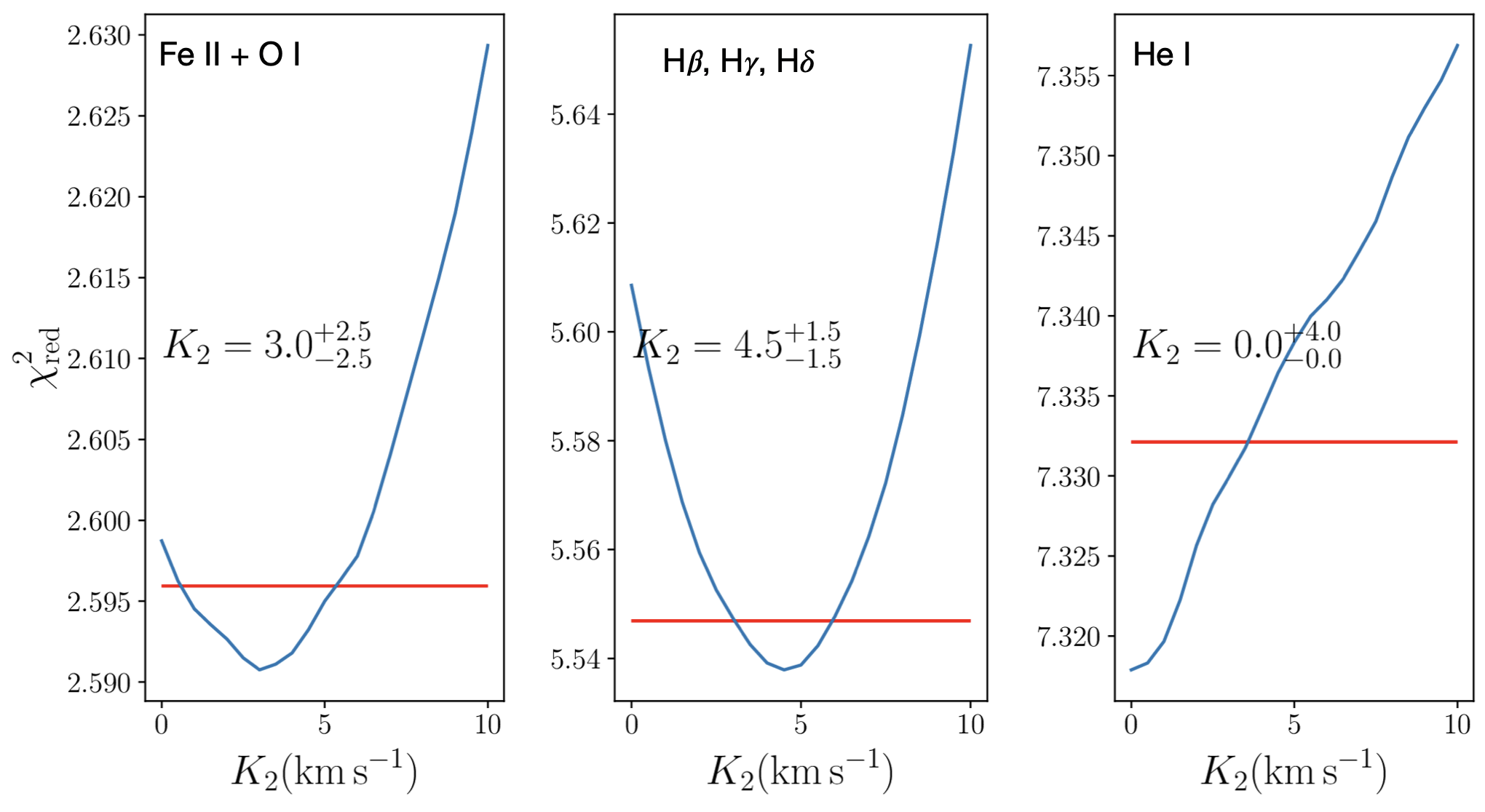}
\caption{Reduced $\chi^2(K_2)$ and corresponding $K_2$ measurements for individual line groups obtained from the observed data set, but using the Fourier spectral disentangling code \texttt{fd3}. The reduced $\chi^2$ is computed in Fourier space and can thus not be directly compared with the one computed by the shift-and-add technique (e.g., Fig.~\ref{fig:MasterGridDis_Main}).
%We observe hugely different optimal $K_2$ values as with the shift-and-add method, as well as, generally, large to extremely large reduced chi-squared values. 
}
\label{fig:fd3}
\end{figure}

Given the low RV amplitude derived for the Be component using the Balmer and Fe\,\textsc{ii} lines, $K_2 = 4\,$\kms{}, an given the discrepant results for the He\,\textsc{i} lines, we investigate the robustness of the disentangling process using artificial data. For this purpose, we create 51 mock spectra of a binary providing the two disentangled spectra as an input, and % applying the same resolving power and 
applying an average S/N comparable to the one in the FEROS data. The mock data mimic the 2004 epoch in terms of their baseline coverage. We fix the orbital parameters of the primary to those given in Table\,\ref{tab:Bstar_properties}, but perform two different tests for two distinct $K_{\rm 2, true}$ values: $K_{\rm 2, true} = 0$\,\kms{} (static Be) and $K_{\rm 2, true} = 4\,$\kms{} (the result inferred here). 

We then perform shift-and-add and Fourier disentangling along the $K_2$-axis on each of these mock data sets. We use the same lines that were used in our study (see Table\,\ref{tab:K2Meas}). As performed for our analysis, we construct reduced $\chi^2$ plots for groups of lines, and exclude the \halpha{} line.

\begin{figure}[t]
\centering
\begin{tabular}{c}
     \includegraphics[width=0.5\textwidth]{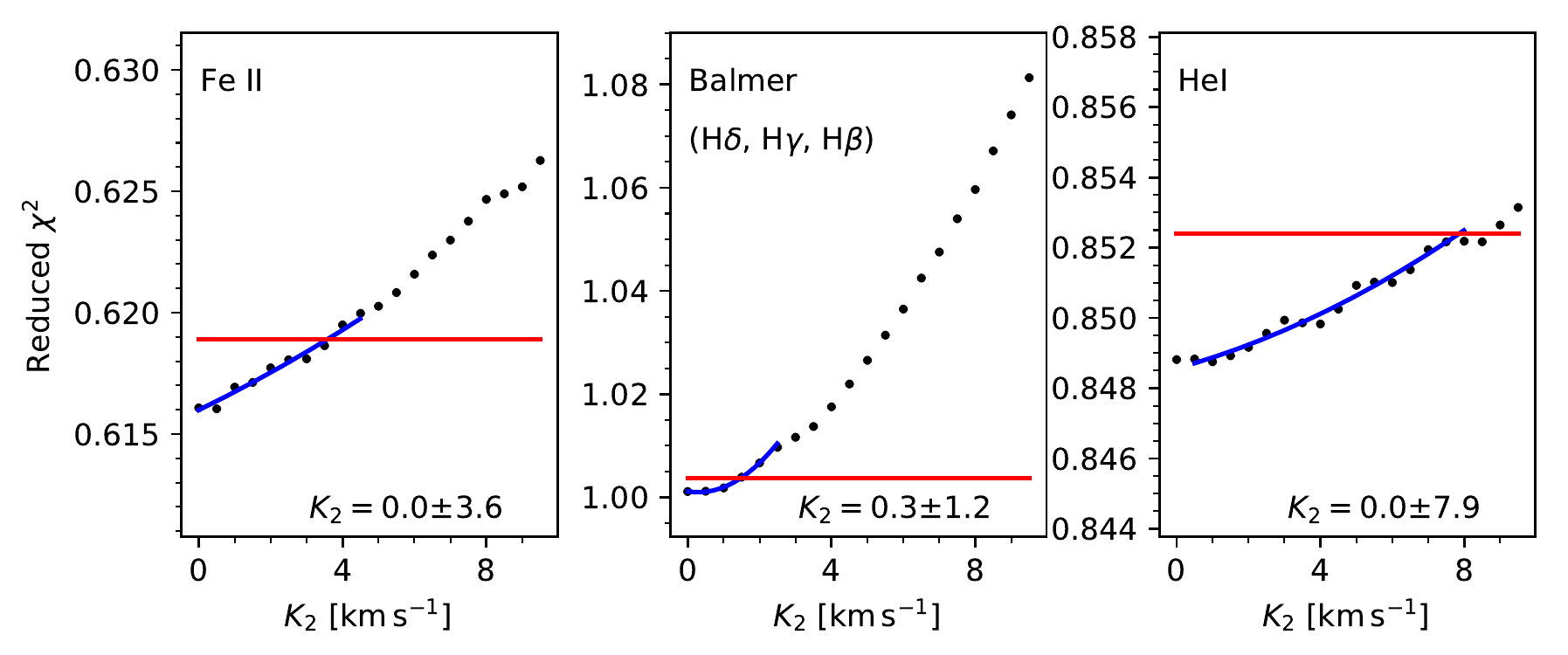} \\
     \includegraphics[width=0.5\textwidth]{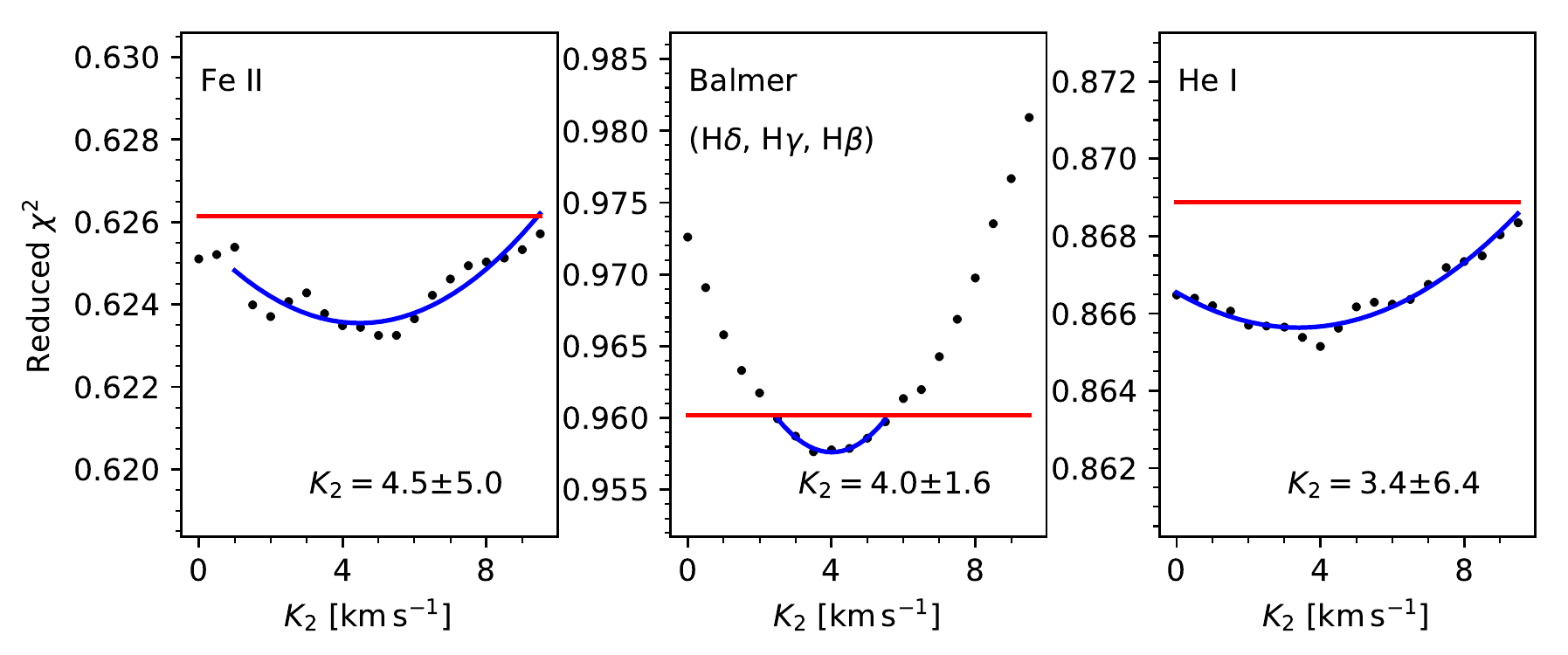} 
\end{tabular}
\caption{Reduced $\chi^2(K_2)$ values obtained for 51 simulated spectra with  $K_{\rm 2, true}=0$ (top) and $K_{\rm 2, true}=4$ (bottom), illustrating that the shift-and-add disentangling is successful in retrieving the true $K_2$ values well within measurement errors.}
\label{fig:SimTestSaA}
\end{figure}

The results for the shift-and-add technique are shown in Fig.\,\ref{fig:SimTestSaA}.  Evidently, all three line-groups are successful in retrieving the true $K_{\rm 2, true}$ values well within their respective 1$\sigma$ errors. The Balmer lines do so most accurately, owing to their high S/N and sharp contrasts. This experiment implies that our results are meaningful. However, we cannot rule out that non-Keplerian variability introduces an additional source of error that is not treated here.

\begin{figure}[t]
\centering
\includegraphics[width=0.5\textwidth]{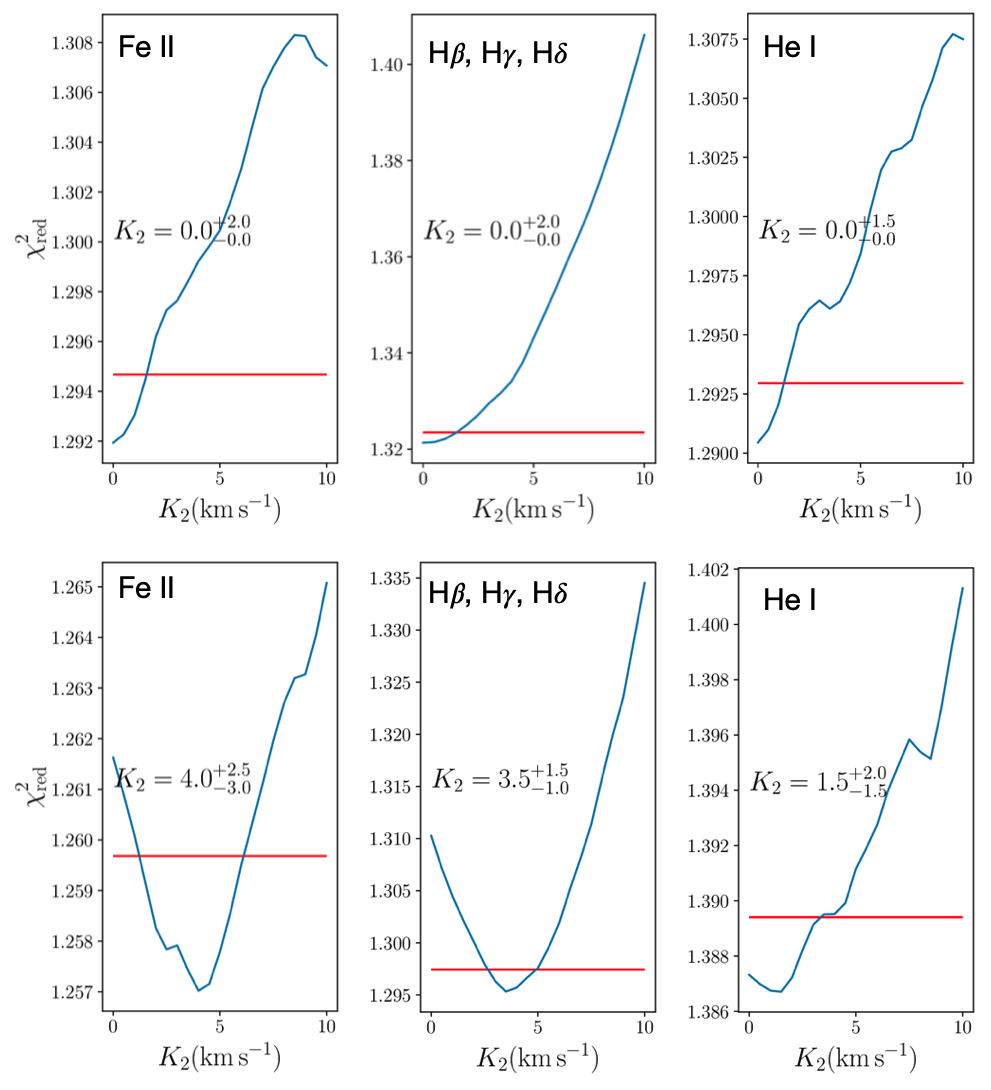}
\caption{Same as Fig.\,\ref{fig:SimTestSaA}, but using Fourier grid disentangling with \texttt{fd3} (top: $K_{\rm 2, true}=0$\,\kms{}, bottom: $K_{\rm 2, true}=4$\,\kms{}), illustrating the discrepant results obtained using Fourier disentangling in this case. The $K_2=4$\,\kms{} semi-amplitude is not unanimously resolved with the Fourier method as it was with the shift-and-add method, even though the reduced chi-squared statistic is of order one.}
\label{fig:sim_fd3}
\end{figure}

We perform the same experiment with the Fourier disentangling code \texttt{fd3} (see Fig. \ref{fig:sim_fd3}). We find that the Balmer and Fe\,\textsc{ii} lines retrieve the input values of $K_2$. However, in the $K_{2,\mathrm{true}}=4$\,\kms{} case, the He\,\textsc{i} lines yields significantly lower values than the input value. These experiments further show that the \ion{He}{i}-based disentangling can hardly be used to distinguish between the cases with  $K_{2,\mathrm{true}}=0$ and 4\,\kms{} given the large errors and systematic bias observed.

%This simple simulation shows that the Fourier disentangling method has its shortcomings finding low amplitude components, at least in this system. A more systematic investigation of this behavior is required to explain why it is happening, but we conjecture that it results from the interplay between the line shapes, the difference in rotation rates, the extreme mass ratio and the stellar variability.


% \section{$K_2$ measurements of individual lines}\label{sec:indilines}
% In Figures \ref{fig:MasterGridDis_Fe}, \ref{fig:MasterGridDis_Balmer}, and \ref{fig:MasterGridDis_He} we show the reduced $\chi^2$ as a function of $K_2$ for individual lines, as measured for the 2004 epoch. Evidently, the different measurements are consistent with each other within $1\sigma$. $K_2$ values larger than 0\,\kms{} are preferred for the vast majority of measurements.

% \begin{figure}
% \centering
% \includegraphics[width=0.5\textwidth]{MasterGrid_Fe.pdf}
% \caption{Reduced $\chi^2(K_2)$ obtained from disentangling the observed spectra of \Starname{}, focusing on various Fe\,{\sc ii} and O\,{\sc i} emission lines. The red lines depict the 1$\sigma$ confidence interval, while the blue curves show the parabola fit to the minima. The derived $K_2$ values along with their 1$\sigma$ errors are given.}
% \label{fig:MasterGridDis_Fe}
% \end{figure}

% \begin{figure}
% \centering
% \includegraphics[width=0.5\textwidth]{MasterGrid_Balmer.pdf}
% \caption{Like Fig.\,\ref{fig:MasterGridDis_Fe}, but for the Balmer lines. }
% \label{fig:MasterGridDis_Balmer}
% \end{figure}

% \begin{figure}
% \centering
% \includegraphics[width=0.5\textwidth]{MasterGrid_He.pdf}
% \caption{Like Fig.\,\ref{fig:MasterGridDis_Fe}, but for several He\,{\sc i} lines. }
% \label{fig:MasterGridDis_He}
% \end{figure}

\section{List of potential progenitor systems}\label{sec:sb9list}% from the $S_{B^9}$ catalogue}\label{sec:sb9}
Table \ref{tab:sb9} lists all systems from the $S_{B^9}$ catalogue that are used on the discussion of the evolutionary scenario of \Starname{} in Section \ref{subsec:somethingsomethingevolution}.

\begin{table*}[h]
 \caption[]{List of binaries considered from the $S_{B^9}$ catalogue. Mass ratios and periods are taken from the catalogue, while the classification as detached, semi-detached or contact comes from the specified reference.}
 \label{tab:sb9}
 \begin{center}
\begin{tabular}{lccc}
 \hline \hline
  Name &
  $q$ &
 Period [d]&
  Reference
 \\ \hline
\noalign{\smallskip}
\multicolumn{4}{c}{Detached} \\
\noalign{\smallskip}
\hline
HS Her & 0.3900 & 1.637 & \cite{KhaliullinKhaliullina2006}\\
$\zeta$ Phe   & 0.6491 & 1.667 & \citet{Andersen1983}\\
VV Ori    & 0.4503& 1.485 & \citet{Terell+2007}\\
IQ Per   & 0.4932 & 1.744 & \citet{LacyFrueh1985}\\
LT CMa& 0.6016 & 1.760  & \citet{Bakis+2010}\\
IM Mon& 0.6062 & 1.190  & \citet{Bakis+2011}\\
$\mu_1$ Sco& 0.6451 & 1.446  & \citet{vanantwerpenMoon2010}\\
V526 Sgr& 0.7400 & 1.920  & \citet{LacySandberg1997}\\
CO Lac & 0.8228 & 1.542 & \citet{Svaricek+2008}\\
AH Cep & 0.8286 & 1.775 & \citet{Holmgreen+1990}\\
RX Her & 0.8466 & 1.779 & \citet{Jeffreys1980}\\
V470 Cyg & 0.8775 & 1.874 & \citet{Russo+1982}\\
U Oph & 0.9027 & 1.677 & \citet{Vaz+2007}\\
V760 Sco & 0.9277 & 1.731 & \citet{Andersen+1985}\\
DW Car   & 0.9389 & 1.328 & \citet{Clausen+2007}\\
AO Mon   & 0.9487 & 1.885 & \citet{Giuricin+1980}\\
PV Cas  & 0.9791& 1.751 & \citet{Popper1987}\\
FZ Cma  & 0.9908 & 1.273 & \citet{GiuricinMardirossian1981} \\
\hline
\noalign{\smallskip}
\multicolumn{4}{c}{Semi-detached} \\
\noalign{\smallskip}
\hline
V1425 Cyg & 1.558 & 1.252 & \citet{Degirmenci+1996}\\
AI Cru   & 1.635 & 1.418 & \citet{Bell+1987}\\
IU Aur & 1.993 & 1.812 & \citet{Harries+1998}\\
V Pup & 2.035 & 1.455 & \citet{Andersen+1983}\\
V1898 Cyg & 5.210 & 1.513 & \citet{Dervisoglu+2011}\\
\hline
\noalign{\smallskip}
\multicolumn{4}{c}{Contact or near-contact} \\
\noalign{\smallskip}
\hline
V606 Cen   & 0.5270 & 1.495 & \citet{Lorenz+1999}\\
SX Aur   & 0.5418 & 1.210 & \citet{Ozturk+2014}\\
$\pi$ Sco   & 0.6328 & 1.570 & \citet{Lefevre+2009}\\
SV Cen   & 0.7075 & 1.659 & \citet{Drechsel+1982}\\
RZ Pyx  & 0.8211 & 0.6563 & \citet{Ergang+2018}\\
GO Cyg   & 0.8528 & 0.7178 & \citet{Ulas+2012}\\
V701 Sco   & 0.9931 & 0.7619 & \citet{BellMalcolm1987}\\
\hline
\noalign{\smallskip}
\multicolumn{4}{c}{Unknown} \\
\noalign{\smallskip}
\hline
HD 23625   & 0.7193 & 1.941 & \\
HD 175544   & 0.7855 & 1.986 & \\
V599 Aql    & 0.6244 & 1.849 & \\
$\sigma$ Aql & 0.7894 & 1.950 & \\
HD 191566 & 0.5985 & 1.818 & \\
7 CrB A & 0.9043 & 1.724 & \\
HD 222900 & 0.8789 & 0.6847 & \\
\hline
\end{tabular}
%\tablebib{(1)~\citet{Andersen1983};
%(2) \citet{xx};
%}
\end{center}
\end{table*}

\section{Setup of MESA simulations}\label{sec:mesa}
The simulation presented in Section \ref{subsec:somethingsomethingevolution} was computed with version 12778 of the MESA code, using an initial composition of $X=0.7$, $Y=0.28$ and $Z=0.02$ for both stars. For simplicity, we do not include stellar rotation, or tidal coupling in this simulation. Opacities are computed from the OPAL project \citep{IglesiasRogers1996}, using solar-scaled metal abundances from \citet{Grevesse+1996}. We use the OPAL equation of state \citep{RogersNayfonov2002}. Nuclear reaction rates are taken fron \citet{Angulo+1999} and \citet{Cyburt+2010}, and we use the 8-isotope network \texttt{basic.net} which includes H$^1$, He$^3$, He$^4$, C$^{12}$, N$^{14}$, O$^{16}$, Ne$^{20}$ and Mg$^{24}$ \citep{Paxton+2011}. Convective boundaries are determined using the Ledoux criterion, with exponential overshooting given by parameters $f=0.01$ and $f_0=0.005$. Semiconvective mixing is modelled as in \citet{Langer+1983} with an efficiency parameter $\alpha_{\rm sc}=1$, and thermohaline mixing follows the model of \citet{Kippenhahn+1980} with an efficiency parameter $\alpha_{\rm th}=1$. We do not include mass loss from stellar winds. Mass transfer due to Roche lobe overflow is computed as in \citet{KolbRitter1990}. Although the model experiences a short-lived contact phase during fast Case A mass transfer, for simplicity we ignore this and apply the model of \citet{KolbRitter1990} as if the system remained semi-detached. A complete set of input files to reproduce our simulation will be included upon acceptance of the paper.
\end{document}